\documentclass[twocolumn,showpacs,amsmath,amssymb,prl,superscriptaddress]{revtex4}
\usepackage{graphicx}
\usepackage{bm}\usepackage{dcolumn}
\usepackage{color}
\newcommand{\sign}{\mathrm{sign}}

\begin{document}
\title{Persisting Meissner state and incommensurate phases \\
 of hard-core boson ladders in a flux}
\author{M. Di Dio}
\affiliation{CNR-IOM-Democritos National Simulation Centre, UDS Via Bonomea 265, I-34136, Trieste, Italy}
\author{S. De Palo}
\affiliation{CNR-IOM-Democritos National Simulation Centre, UDS Via Bonomea 265, I-34136, Trieste, Italy}
\affiliation{Dipartimento di Fisica Teorica, Universit\`a di Trieste, Trieste, Italy}
\author{E. Orignac}
\affiliation{Laboratoire de Physique de l'\'Ecole Normale Sup\'erieure de Lyon,
CNRS UMR5672, 46 All\'ee d'Italie, F-69364 Lyon Cedex 7, France}
\author{R. Citro}
\affiliation{Dipartimento di Fisica "E.R. Caianiello", Universit\`a
  degli Studi di Salerno and Unit\`a Spin-CNR , Via Giovanni Paolo II, 132, I-84084 Fisciano
  (Sa), Italy}
\author{M.-L. Chiofalo}
\affiliation{Dept. of Physics "Enrico Fermi" and INFN,  Universit\`a di
  Pisa Largo Bruno Pontecorvo 3 I-56127 Pisa, Italy}
\date{\today}
\begin{abstract}
  The phase diagram of a  half-filled  hard core boson two-leg ladder
  in a flux is investigated by means of numerical simulations based on the Density Matrix
  Renormalization Group (DMRG)  algorithm and  bosonization.
 We calculate experimentally accessible observables such as the momentum distribution, as well as rung current, density
  wave and bond-order wave correlation functions, allowing us to
  identify the Mott Meissner and Mott Vortex states.
 We
  follow the transition from commensurate Meissner to incommensurate
  Vortex state at increasing interchain hopping till the critical
  value [Piraud \textit{et al.} Phys. Rev. B \textbf{91}, 140406 (2015)] above which the  Meissner state is stable at any flux.
  For flux close to $\pi$, and below the critical hopping,  we observe
  the formation of a   second  incommensuration  in the Mott
  Vortex state that could be detectable in current experiments.
\end{abstract}
\pacs{03.75.Lm,05.30.Rt,64.70.Rh,71.10.Pm}
\maketitle
Superconductors in external magnetic field $H<H_{c1}$ exhibit the
Meissner-Ochsenfeld effect where surface currents
screen completely the magnetic field in the bulk, resulting in
perfect diamagnetism~\cite{tinkham_book_superconductors}. Type-I
superconductors return to the normal state for $H>H_{c1}$ while
in type-II superconductors, for
$H_{c1}<H<H_{c2}$ a vortex phase is formed, in which the magnetic
field partially penetrates the system along flux lines surrounded by
screening currents. This behavior can be understood in the framework
of spontaneous breaking of a global U(1) symmetry via the
Landau-Ginzburg equation\cite{tinkham_book_superconductors}.
In a quasi-one dimensional system, such symmetry breaking is precluded
by the Mermin-Wagner-Hohenberg
theorem\cite{mermin_wagner_theorem,hohenberg67_theorem}. However, in
the case of a bosonic two-leg ladder~\cite{kardar_josephson_ladder,orignac01_meissner,cha2011},
an analog of the Meissner phase was predicted to exist in the ground
state for low flux, while for higher flux a Tomonaga-Luttinger liquid
(TLL) of vortices was expected. The quantum phase transition between these
two states is in the commensurate-incommensurate (C-IC) universality class
\cite{japaridze_cic_transition,pokrovsky_talapov_prl}.
Other orderings have been predicted, such as chiral superfluid order at half a flux
quantum per plaquette~\cite{kardar_josephson_ladder,granato_josephson_ladder,nishiyama_josephson_ladder} and
a chiral Mott insulating
phase\cite{dhar2012,dhar2013,petrescu2013,tokuno2014,petrescu2015},
which is a Mott regime\cite{keles2015} possessing chiral currents as well as a
spin-density wave phase. 
DMRG
studies of ladders with diagonal interchain hopping are also available~\cite{zhao2014a,zhao2014b,xu2014,peotta2014,piraud2014,barbiero2014}.
While the original proposal was made in the context of Josephson junction
ladders, where the quantum effects are spoiled by
 dissipation~\cite{fazio_josephson_junction_review}, the advent of
 ultracold atomic gases offers another realization of strongly
 interacting one dimensional boson
 systems\cite{paredes_toks_experiment,kinoshita_tonks_experiment}.
Moreover, it has been shown
theoretically~\cite{osterloh05_gauge,ruseckas05_gauge} and
experimentally~\cite{lin2011_soc}  how artificial gauge field
could be created in these systems.   Recently, the transition from Meissner to
Vortex phases in non-interacting bosonic ladders of ultracold atoms has been
studied experimentally at fixed flux $\pi/2$  per plaquette and
variable interleg hopping~\cite{atala2014}.

\begin{figure}[h]
\begin{center}
\includegraphics[trim = 12mm 5mm 5mm 5mm,width=82.mm]{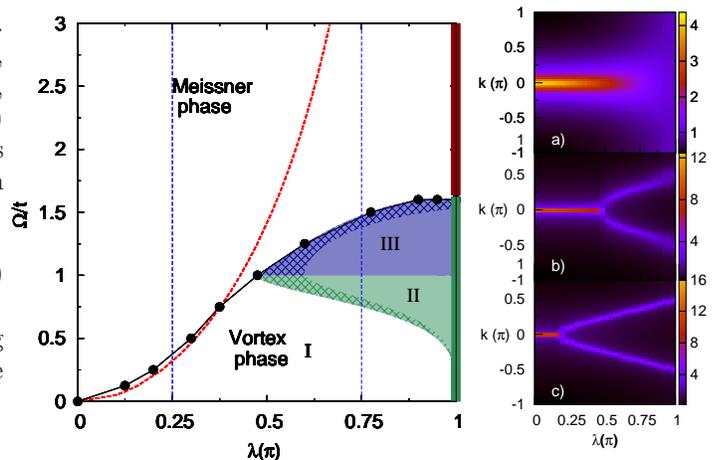}
\end{center}
\caption{(color online) Left panel: Phase diagram of (\ref{eq:H-model}) as a function of flux per
plaquette $\lambda$ and $\Omega/t$. The phase boundary between the Meissner and the Vortex phase
is shown by the black solid line, displaying the persistence of the
Meissner phase\cite{piraud2014b} above the threshold $\Omega>\Omega_c$ for all fluxes $\lambda$, except $\lambda=\pi$. For comparison the red-dashed line shows the boundary $\lambda^{(0)}_c(\Omega)$
for the non-interacting system. In the shaded area a second incommensuration appears.
In the green area (region II) extra peaks at $k=\pi$ develop in
the FT of the rung current correlations and they become the dominant
correlations in the blue region (III). The double line (green vs dark red) at $\lambda=\pi$
represents the transition to a localized phase.
In the right panel we show intensity plots of $n(k)$ versus $\lambda$ and $k$. In panel (a), at $\Omega/t=1.75$
the system is always in the MM phase, indeed only a single maximum at $k=0$ is visible for all
$\lambda$. At $\lambda=\pi$, $n(k)=1$ indicating the formation of a fully
localized state  (dark red solid line).
In panel (b) and (c), for $\Omega/t=1$ and $0.25$ respectively, the transition from
MM to the MV with two maxima symmetric around $k=0$, is shown.}
\label{fig:phase_diagram}
\end{figure}

In this Letter we explore the phase diagram of  hard-core spinless
bosons on a two-leg ladder at  half-filling  as a function of flux and
interchain hopping by means of numerical simulations using DMRG algorithm and  bosonization.
We find, in agreement with \cite{piraud2014b}, that hard-core constraints cause a significant enlargement of the Meissner phase over the vortex one with respect to the non--interacting case: above a critical value of the interchain
hopping\cite{piraud2014b} the system
remains in the Mott-Meissner (MM) state for any flux (see Fig.~\ref{fig:phase_diagram}).
Below the critical interchain hopping, both the behavior of the momentum distribution and of the rung current, show that the transition from Mott-Meissner (MM) to Mott-Vortex (MV) state falls in the universality
class of the C-IC
transition\cite{orignac01_meissner}. For  fluxes close to
$\pi$, we observe another incommensuration, whose origin is
discussed within bosonization.

We consider\cite{orignac01_meissner} a two-component system of hard core bosons on two leg
ladder, with a flux per plaquette
$\lambda$ and interchain hopping $\Omega$: \begin{eqnarray}
H_\lambda &=&
-t\sum_{j,\sigma}\left(b^\dagger_{j,\sigma}e^{i\lambda\sigma}b_{j+1,\sigma}+
  \mathrm{H. c.}\right) \nonumber \\
&+&
\Omega\sum_j\left(b^\dagger_{j,\uparrow}b_{j,\downarrow}+\mathrm{H. c.}\right),
\label{eq:H-model}
\end{eqnarray}
with $b^\dagger_{j,\sigma}(b_{j,\sigma})$ bosonic creation
(annihilation) operator at site $j$, $\sigma=\pm 1/2$ the chain index,
 and $t e^{i\lambda\sigma}$ the hopping
amplitude along the chain $\sigma$.
This Hamiltonian can be mapped onto a system of spin-1/2~\cite{cazalilla2011} bosons with
spin-orbit coupling in a transverse magnetic field ~\cite{supplementary}
with each  spinor state  corresponding to one leg of the ladder.
For half-filling, i.e. for one boson per rung, at $\lambda=0$ and $\Omega \ne 0$ the
ground state of (\ref{eq:H-model}) is a rung-Mott Insulator\cite{crepin2011}.
For $\lambda>0$, according to the bosonization
treatment\cite{haldane_bosons},
two phases with a charge gap are expected\cite{orignac01_meissner,petrescu2013,tokuno2014,petrescu2015}, the Mott-Meissner (MM) and the Mott-Vortex (MV) state.
In the MM state, for $0<\lambda<\lambda_c$,  two currents of opposite sign flow along the legs\cite{piraud2014b},
the interchain current
\begin{equation}
J_r(l) =i\Omega \left( b^\dagger_{l,\uparrow}b_{l,\downarrow}-b^\dagger_{l,\downarrow}b_{l,\uparrow},
 \right)
\end{equation} has zero expectation value and exponentially decaying correlations,  and
the screening current, {\it i.e} the difference between the currents
of the two legs
\begin{eqnarray}
J_s=-it\sum_{j,\sigma}\left(\sigma
e^{i\lambda\sigma}  b^\dagger_{j,\sigma}b_{j+1,\sigma} -\sigma
e^{-i\lambda\sigma} b^\dagger_{j+1,\sigma}b_{j,\sigma} \right),
\label{eq:Js}
\end{eqnarray}
is a smooth function of the applied flux (increasing linearly at small flux).  On
increasing the flux $\lambda>\lambda_c(\Omega)$, the system enters  the MV state, there is a sudden
drop\cite{piraud2014b} of the screening current $J_s$ and simultaneously the rung
current correlations decay becomes algebraic\cite{piraud2014b} with  an
incommensurate modulation of wavevector $q(\lambda)$.  Close to the
transition point $\lambda_c(\Omega)$, the wavevector
$q(\lambda)\sim \sqrt{\lambda^2-\lambda_c^2}$.
In the non-interacting case, the Hamiltonian Eq.~(\ref{eq:H-model}) can
be readily diagonalized\cite{tokuno2014,keles2015} and
$\lambda^{(0)}_c(\Omega)=2\arctan[\Omega/(4t)]$.
The occurrence of the MV phase can be seen out also in the total, as well as,
in the spin resolved momentum distribution\cite{cha2011} of the system:
\begin{equation}
n(k)=\sum_\sigma n_\sigma(k)=\frac{1}{L}\sum_\sigma \sum^{L-1}_{i,j} e^{i k (r_i-r_j)}
\langle b^\dagger_{i,\sigma} b_{j,\sigma} \rangle.
\label{eq:nk}
\end{equation}
In the MM phase $n(k)$ has a single maximum at $k=0$, whereas in the MV phase it exhibits
a pair of maxima $k=\pm q(\lambda)/2$\footnote{Note that the momentum distribution is a
gauge dependent quantity, and that our gauge choice differs from the
one of \cite{cha2011}.}.
We have obtained the ground state phase-diagram of (\ref{eq:H-model})
by computing various observables  like the momentum distribution and the screening
current $J_s$ together with the Fourier Transform (FT)
$C(k) = \sum_l e^{-ik l} \langle J_r(l) J_r(0)\rangle$ of the rung current correlation function.\\
While performing simulations with both periodic (PBC) and open (OBC) boundary conditions,
we found the former to be more suitable
for our system, despite the well-known computationally more demanding
convergence properties typical of PBC~\cite{white_dmrg_letter,white_dmrg,schollwock2005}.
As such we run simulations employing PBC for system sizes ranging from $L=16$ to $L=64$, keeping up to
$m=1256$ states during the renormalization procedure. In this way the truncation error {\it i.e.} the weight of
the discarded states, is at most of order $10^{-6}$, while the maximum error on the ground-state energy is of
order $5\times10^{-5}$ at its most.
We further extrapolate in the limit $m \rightarrow \infty$ all the quantities calculated to
characterize the phase diagram.
In Fig.~\ref{fig:phase_diagram}, we summarize our findings for the phase diagram
at half-filling. At variance with the non-interacting case where there
is a critical $\lambda^{(0)}_c(\Omega)$ for all $\Omega$, in
 the presence of the hard-core interaction, for interchain hoppings
 $\Omega>\Omega_c$, the commensurate-incommensurate transition
 disappears\cite{piraud2014b} and  the MM phase
is stable for all fluxes. Another effect of the hard-core interaction, as we will
discuss below, is that in the Vortex phase, at $\lambda=\pi$ and $\lambda$ close to $\pi$,
a commensurate peak appears in $C(k\simeq \pi)$, along with an
incommensuration in  the density correlations. At $\lambda=\pi$, and
for  $\Omega>\Omega_c$  a fully rung localized phase is obtained. Such
rung localized ground state was discussed in the limit $\Omega\gg t$
in \cite{piraud2014b}. 

We have characterized the nature of the Mott-Meissner and Mott-Vortex phases
by examining $C(k)$, the staggered boson density wave $S(k)$  and the
symmetric bond--order wave $S^c_{BOW}$ static structure factors which bring information on
the spin density and bond-order waves, respectively:
\begin{eqnarray}
\label{eq:Ss}
&& S(k)=\frac 1 L \sum_{j,l=0 \atop \sigma \sigma' }^{L-1}
 e^{ik(j-l)}\mathrm{sign}(\sigma\sigma') \langle  n_{j,\sigma} n_{l\sigma'} \rangle, \\
\label{eq:Sbc}
&& S^c_{BOW}(k)=\frac 1 L \sum_{j,l=0}^{L-1}
 e^{ik(j-l)}\langle \delta B_{j}\delta B_{l}\rangle;
\end{eqnarray}
where $B_j =\sum_{\sigma} b^\dagger_{j+1,\sigma}b_{j,\sigma}+ H.c.$
and $\delta B_j=B_j-\langle B_j \rangle $.

\begin{figure}[h]
\begin{center}
\includegraphics[height=65.5mm]{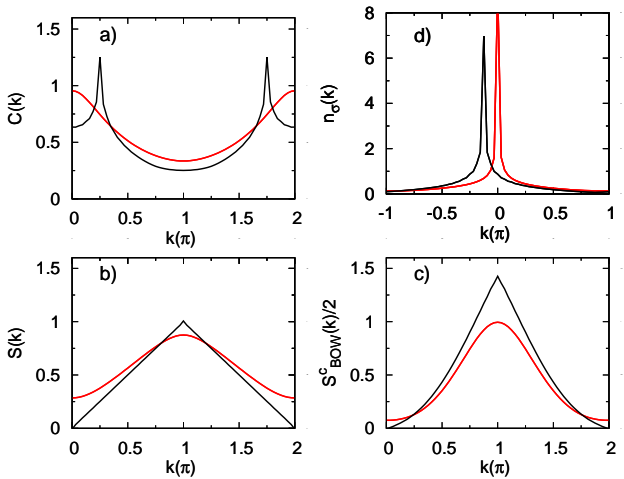}
\end{center}
\caption{We show FT of correlation functions as from DMRG simulation for $L=64$
at $\lambda=\pi/4$ for two different values of the $\Omega/t=0.0625$ and $1$, respectively in the Vortex
(black solid line) and Meissner phase( red solid line). Panel a) shows the FT of the rung-current correlation
function $C(k)$, panel b) the spin correlation functions $S(k)$ and panel c) the charge bond-order correlation function $S^c_{BOW}$. In
panel d)  the spin resolved momentum distribution is shown, with $n_{-\sigma}(k)=n_{\sigma}(-k)$.}
\label{fig:small}
\end{figure}

In Fig.~\ref{fig:small} we follow the the MM--MV phase transition at small
$\lambda$ and $\Omega$ (see cut one in Fig.1). As predicted from bosonization
~\cite{supplementary} the vortex phase is signalled by the
appearance in $C(k)$
of two cusp-like peaks respectively at $k=q(\lambda)$ and $k=2\pi-q(\lambda)$
(see panel a) of Fig.~\ref{fig:small} whose heights do not scale with the size of the
system (see Fig.~1 of \cite{supplementary}).
In MV phase, the spin resolved momentum distribution $n_{\sigma}(k)$ shows a symmetric
peak centered at $k=\sigma q(\lambda)$, as predicted by
bosonization. In this region of parameter space
the correlation length associated with the Mott gap\cite{crepin2011} is comparable with
the system size, and the peak takes a cusp-like shape as in a
Tomonaga-Luttinger liquid\cite{cazalilla2011}, instead of the typical Lorentzian-shape
expected for a Mott-insulator.
Also $S(k)$ shows the expected low momentum behaviour according to bosonization approach:
in the MM phase $S(k) = S(0)+ a k^2 +o(k^2)$, with $S(0)>0$, while in the MV phase
$S(k) =\frac{K_s^* |k|}{\pi} +o(k)$, with $K^*_s=1$ (as expected for a
hard-core boson system) a signature of a TLL of vortices.
 The transition is also seen in $S(k \simeq \pi)$. In the MM
 phase, $S(k\simeq \pi)$ shows a Lorentzian-shaped peak while in the
 MV phase this peak takes a cusp-like shape.
A similar change across the MM-MV transition  is also seen  in
the correlation function
$S_{BOWc}(k\simeq \pi)$(see Supplemental Material\cite{supplementary}).
\begin{figure}[h]
\begin{center}
\includegraphics[height=65mm]{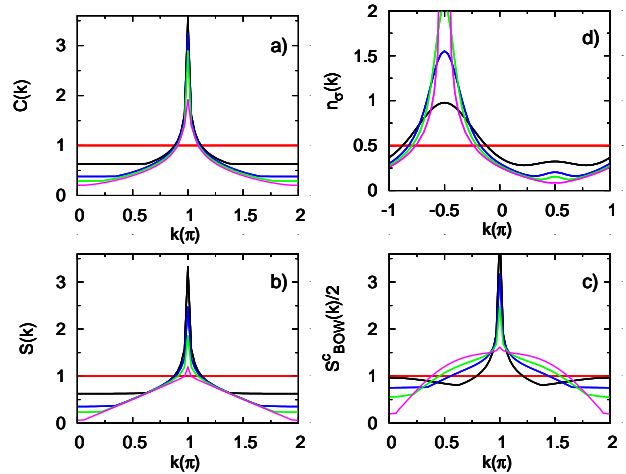}
\end{center}
\caption{We show FT of correlation functions as from DMRG simulations for $L=64$
at $\lambda=\pi$ at various $\Omega/t$. Panel a) shows the FT of the rung-current correlation
function $C(k)$, panel b) the spin correlation function $S(k)$ and panel c) the charge bond-order correlation function $S^c_{BOW}$. In
panel d)  the spin resolved momentum distribution is shown. Red solid curves
are for $\Omega/t=1.75$ in the fully localised state, while black, blue, green and magenta solid lines are respectively
for $\Omega/t=1.5, 1.25,1$ and $0.5$.}
\label{fig:pi}
\end{figure}
This description breaks down when $\lambda$ is no longer a small
quantity as $q(\lambda)$ would be comparable to the momentum cutoff.\\
At $\lambda=\pi$ the major changes from the conventional C-IC transition at small flux are observed.
To derive the low energy Hamiltonian it becomes
necessary to choose
the gauge with the vector potential along the rungs of the ladder, so that
the interchain hopping reads:
\begin{equation}
  \label{eq:gauge-rung}
  H_{hop.}=\Omega \sum_{j,\sigma} (-)^j b^\dagger_{j,\sigma} b_{j,-\sigma}.
\end{equation}
After applying bosonization, the hopping Hamiltonian can be rewritten
in terms of a free boson $\phi_c$ describing the total density
fluctuations coupled  to  $\mathrm{SU(2)}_1$
Wess-Zumino-Novikov-Witten (WZNW) currents $\mathbf{J}_{R,L}$
describing the chain antisymmetric density fluctuations by a term
$\propto \Omega \cos \sqrt{2} \phi_c (J_R^y + J_L^y)$  ( see
~\cite{supplementary} for details). Such a term can be
treated in mean-field
theory\cite{nersesyan_incom,lecheminant2001,jolicoeur2002,zarea04_chiral_xxz}. This
procedure leads to an effective Hamiltonian with a gap $\Delta_c\sim
\Omega^2$ for the total density excitations, while the
antisymmetric density modes remain gapless and develop an
incommensuration of wavevector $p(\Omega)\propto \Omega^2$
(see Fig.~2 in \cite{supplementary}).
The presence of this predicted incommensuration is visible in the low momentum behaviour of
$S(k)$ and $C(k)$ (panel a) and panel b) of Fig.~\ref{fig:pi} that become $\propto
\frac{K_s^*}{2\pi}(|k-p(\Omega)|+|k+p(\Omega)|)$, {\it i.e.} constant
for $|k|< p(\Omega)$ and linear in $k$ for $|k|> p(\Omega)$.
In the $S^c_{BOW}(k)$ we observe a cusp at the same vectors
$p(\Omega)$ (panel c) of Fig.~\ref{fig:pi}). As expected, all these
correlation functions  also develop a peak at $k=\pi/a$.
A sign of the incommensurability at $\lambda=\pi$ should
be visible also in the momentum distribution $n_\sigma(k)$
(see Fig.~\ref{fig:pi}, panel d)). In this case, a calculation based on non-abelian
bosonization and operator product expansion, would lead to three Lorentzian-like peaks
centered in $\pi/(2a)$ and $\pi/(2a) \pm p(\Omega)/2$. However, these peaks cannot be
separated if the correlation length in real space $u_c/\Delta_c \sim
\Omega^{-2}$ is shorter than the wavelength $2\pi/p(\Omega) \sim \Omega^{-2}$. In the
numerical simulations, at $L=64$ in PBC,~(see Fig.~\ref{fig:pi}) a broad peak is
observed for $k=\frac\pi{2a}$.

When $\lambda \alt \pi$ (second cut in Fig.1) we can proceed analogously to the previous case and choose a gauge
such that: \begin{eqnarray}
H&=&-t \sum_{j,\sigma} \left( b^\dagger_{j,\sigma}
 e^{i(\lambda-\pi)\sigma}b_{j+1,\sigma} + \mathrm{H.c.} \right)
\nonumber \\  && + \Omega \sum_{j,\sigma} (-1)^j b^\dagger_{j,\sigma} b_{j,-\sigma},
\end{eqnarray}
and define $\delta \lambda = (\lambda -\pi)$, so that the bosonized
Hamiltonian contains the extra term $\delta \lambda (J_R^z -J_L^z)$.
For this case, the Fourier transform of the rung current correlation will present peaks
at $k=\frac \pi a$ and $k=\frac \pi a  \pm \sqrt{p(\Omega)^2 + (\delta \lambda/a)^2}$.
When $\delta \lambda$ is increased, these last two peaks
become dominant, and we crossover to the behavior already discussed
for weak $\lambda$.
\begin{figure}[h]
\begin{center}
\includegraphics[height=65mm]{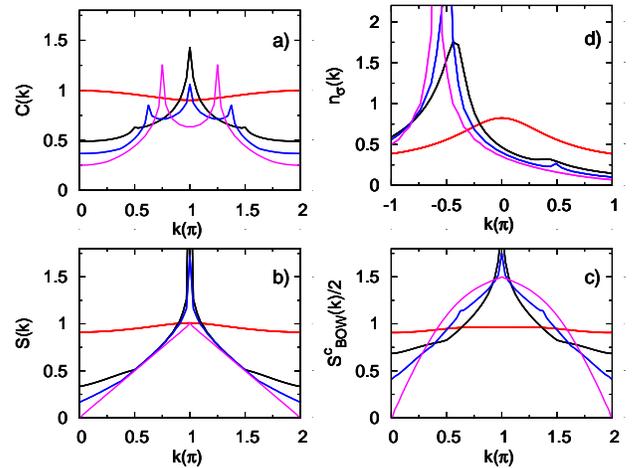}
\end{center}
\caption{We show FT of correlation functions as from DMRG simulations for $L=64$
at $\lambda=3\pi/4$ at various $\Omega/t$. Panel a) shows the FT of the rung-current correlation
function $C(k)$, panel b) the spin correlation function $S(k)$ and panel c) the symmetric bond-order correlation function $S_{BOWc}$. In
panel d)  the chain resolved momentum distribution is shown, with $n_{-\sigma}(k)=n_{\sigma}(-k)$. Red solid curves
are for $\Omega/t=1.5$  in the Meissner phase, while black, blue, green and magenta solid lines are respectively
for $\Omega/t=1.25, 1$ and $0.625$.}
\label{fig:near}
\end{figure}
At $\lambda < \pi$ the $C(k)$ (see Fig.~\ref{fig:near}) shows, beside
the peak at $k=\frac \pi a$,
two peaks symmetric around $k=\frac \pi a$, in real space these last two oscillations exibit an
exponential decay for $\Omega/t > 1$ and a power law for $\Omega/t \leq 1$ (region III and II in Fig.1). The situation is
reversed for the oscillation at $k=\frac\pi a$. At $\Omega/t=1$ all oscillations, for systems with $L=64$
in PBC, exhibit power law decay.
The effect of this incommensuration can also be followed in the behaviour at small $k$
of $S(k)$ that instead of being a constant value for
$k < \sqrt{p(\Omega)^2 + (\delta \lambda/a)^2}$ shows a linear behaviour.
In  $S^c_{BOW}(k)$, for $\Omega/t \leq 1$ two symmetric peaks are
present at $k=\pm q(\lambda)$.
We checked that the phase is a single component Tomonaga-Luttinger liquid
by computing  the Von Neumann entropy at $\Omega/t=1.5$ for $\lambda=0.75\pi$ and $0.8125 \pi$
and obtaining the expected logarithmic dependence with
system size\cite{piraud2014b}, ruling out a a Chiral Mott
insulator\cite{petrescu2013,petrescu2015} for $\lambda \alt \pi$.
At general commensurate filling $n$ and flux $\lambda=2\pi n$,
the incommensuration generating term becomes $-i \Omega
e^{i\sqrt{2}\phi_c} (J_R^++J_L^-) + \mathrm{H. c.}$, leading to
density wave phases with incommensuration.
Let us finish by noting that such incommensuration
is specific of hard core boson systems. 
With less repulsive interactions, the term that gives rise to the
vortex lattice state would be
relevant\cite{petrescu2013,petrescu2015,greschner2015}, while stronger
repulsion would make the term stabilizing the checkerboard density
wave relevant. Adding a nearest neighbor intrachain interaction $V$ to the hard
core repulsion, a first phase transition at $V<0$ will separate the
vortex lattice from the incommensurate state, and a second transition
at $V>0$ will separate the incommensurate state from the density wave
state.

In conclusion, we have studied a two-leg hard core boson ladder in an artificial
gauge field. In contrast to the non-interacting
case,  the vortex phase is suppressed when the interchain hopping
exceeds a threshold value, as found in \cite{piraud2014b}. 
At flux $\pi$ per plaquette and $\Omega/t>1.5$ the ground state
becomes a tensor product of singly occupied rungs, as was expected\cite{piraud2014b} in the
$\Omega/t \to \infty$ limit. For $\Omega/t<1.5$, we have obtained an incommensurate
insulating state similar to
the spin-nematic state of frustrated XXZ spin
chains\cite{nersesyan_incom,lecheminant2001,jolicoeur2002,zarea04_chiral_xxz}.
In the case of a system of weakly coupled ladders, a long range
ordered phase could form in which density wave or rung current would
possess a long range commensurate order, but
exponentially damped incommensurate correlations would still be
present. The presented results could be detectable in current experiments with cold atoms\cite{atala2014} and the evidence of a persisting Meissner state could be relevant for quantum computing purposes in defining a stable flux qbit\cite{devoret2004,franco2005}. \\

\begin{acknowledgments}
We thank F. Ortolani for the DMRG code. Simulations were run at
Universit\`a di Salerno
and Universit\`a di Pisa local computing facilities.
M.D.D. and M.L.C. acknowledge partial support from PRIN-2011 "Collective Quantum Phenomena: from
strongly correlated system to quantum simulators".
\end{acknowledgments}

\newpage 
\widetext 
\begin{center}
\textbf{\large Supplemental Material for ``Persisting Meissner state and incommensurate phases
 of hard-core boson ladders in a flux''}
\end{center}
\setcounter{equation}{0}
\setcounter{figure}{0}
\setcounter{table}{0}
\setcounter{page}{1}
\makeatletter
\renewcommand{\theequation}{S\arabic{equation}}
\renewcommand{\thefigure}{S\arabic{figure}}
\renewcommand{\bibnumfmt}[1]{[S#1]}
\renewcommand{\citenumfont}[1]{S#1}

\begin{center}
\textbf{ \large Supplementary material for ``Incommensurate phases of a hard-core boson two-leg ladder in a flux''}
\end{center}

\section{Correlation functions and observables in bosonization}
\label{sec:boso}

\subsection{Bosonized Hamiltonian and observables}
\label{sec:boso-ham}

Let us first consider a single leg $\sigma$ in the case
of $\Omega=0,\lambda=0$.
Hard core bosons are mapped on non-interacting spinless fermions by the
Jordan-Wigner transformation\cite{jordan_transformation}. These
non-interacting spinless fermions are then
bosonized\cite{giamarchi_book_1d}.
 The bosonized form of the
Hamiltonian $H_\parallel$  reads:
\begin{eqnarray}
  \label{eq:2cll-ham-no-lambda}
  H_\parallel=\sum_\sigma \int \frac{dx}{2\pi} \left[ u K (\pi
      \Pi_\sigma)^2 + \frac{u}{K} (\partial_x\phi_\sigma)^2  \right],
\end{eqnarray}
 where
 $[\phi_\sigma(x),\Pi_\sigma'(x')]=i\delta_{\sigma\sigma'}\delta(x-x')$
 and $\pi \int^x \Pi_\alpha=\theta_\alpha$. in
 Eq.~(\ref{eq:2cll-ham}), $u$ is the velocity of excitations, while
 $K$ is the Tomonaga-Luttinger (TL) parameter. For non-interacting
 spinless fermions, $K=1$ and $u=2t a \sin(\pi n/2)$, where $n=\langle
 n_\uparrow+n_\downarrow \rangle $ is the
 average number of bosons per site and $a$ is the lattice spacing. At
 half-filling (\textit{i. e.} $n=1$) $u=2t$. The boson annihilation
 operators are
 represented as\cite{haldane_bosons}\cite{giamarchi_book_1d}:
 \begin{equation}
   \label{eq:boson-operator}
   b_{j,\sigma} = e^{i \theta_\sigma(ja)}\left[A_0 + \sum_{m\ne 0} A_{m}
   e^{2i m (\phi(ja) -\pi  \langle n_\sigma \rangle x/a)} \right].
 \end{equation}
 In the presence of a vector potential along the legs of the ladder,
the lattice Hamiltonian can be brought back to the case
of $\lambda=0$ by the canonical transformation $b_{j,\sigma}=e^{-i j \lambda \sigma} \bar{b}_{j,\sigma}$.
The bosonization technique can then be applied to the Hamiltonian
written in terms of the $\bar{b}_{j,\sigma}$ bosons. One finds a Hamiltonian of
the form~(\ref{eq:2cll-ham-no-lambda}) with
$\bar{\theta}_\sigma,\bar{\phi}_\sigma$ replacing
$\theta_\sigma,\phi_\sigma$. Using Eq.~(\ref{eq:boson-operator}), one obtains
$\theta(x)=\bar{\theta}(x)-\lambda \sigma x/a$, and
$\phi(x)=\bar{\phi(x)}$ giving the bosonized Hamiltonian
\begin{eqnarray}
  \label{eq:2cll-ham}
  H_\parallel=\sum_\sigma \int \frac{dx}{2\pi} \left[ u K \left(\pi  \Pi_\sigma +\sigma \frac{\lambda}{a}\right)^2 + \frac{u}{K} (\partial_x\phi_\sigma)^2  \right],
\end{eqnarray}
Actually,  it is
convenient to turn to symmetric ($c$) and antisymmetric ($s$)  representation,  $\phi_{c,s}=\frac{\phi_\uparrow \pm \phi_\downarrow}{\sqrt{2}}$, $ %
 \theta_{c,s}=\frac{\theta_\uparrow \pm %
   \theta_\downarrow}{\sqrt{2}}$,
 so that the full Hamiltonian reads:
\begin{eqnarray}
  \label{eq:bosonized}
  && H=H_c+H_s^\lambda \\
  \label{eq:bosonized-c}
  && H_c=\int \frac{dx}{2\pi} \left[ u_c K_c (\pi \Pi_c)^2 +
    +\frac {u_c} {K_c} (\partial_x \phi_c)^2 \right]\\
   \label{eq:bosonized-s}
  && H_s^\lambda =\int \frac{dx}{2\pi} \left[ u_s K_s \left(\pi
      \Pi_s+\frac{\lambda}{a\sqrt{2}}\right)^2
    +\frac {u_s} {K_s} (\partial_x \phi_s)^2 \right]
\end{eqnarray}
where $u_c K_c=u_s K_s = u K$, $u_c/K_c=u_s/K_s=u/K$, and we have used $\sigma=\pm 1/2$.

The boson annihilation operators become:
\begin{equation}
b_{j\sigma} =e^{\frac{i}{\sqrt{2}} (\theta_c + 2 \sigma
  \theta_s) }\left[\sum_m A_{m} e^{i m (\sqrt{2} \phi_c + 2\sigma
    \sqrt{2} \phi_s - 2\pi \langle n_\sigma \rangle x/a)}\right]
\end{equation}

Therefore,
\begin{eqnarray}
  \langle b_{j\sigma} b^\dagger_{0 \sigma} \rangle \sim \langle
  e^{\frac{i}{\sqrt{2}} \theta_c(ja)}  e^{-\frac{i}{\sqrt{2}}
    \theta_c(0)}\rangle  \langle
  e^{\frac{i}{\sqrt{2}} \sigma \theta_s(ja)}
  e^{-\frac{i}{\sqrt{2}}\sigma
    \theta_s(0)}\rangle +\ldots 
\end{eqnarray}
Using instead the bosonized expression for the density $\rho_\sigma = \frac{\langle n_\sigma\rangle }{a} -\partial_x\phi_\sigma + B_1 \sin
  (2\phi_\sigma -2\pi \langle n_\sigma \rangle x/a)$ one obtains the 
the total boson density $\rho_{tot}= \rho_\uparrow +
\rho_\downarrow$ as:
\begin{equation}
  \label{eq:density-tot}
  \rho_{tot}(x)=\frac 1 a -\frac{\sqrt{2}}{\pi} \partial_c \phi_c + 2
  B_0 e^{i \pi \frac x a} \sin \sqrt{2} \phi_c \cos \sqrt{2} \phi_s,
\end{equation}
and the antisymmetric density $\sigma^z= (\rho_\uparrow -\rho_\downarrow)/2$  as:
\begin{equation}
  \label{eq:spin-density}
  \sigma^z(x)=-\frac 1 {\pi \sqrt{2}} \partial_x \phi_s + B_0 e^{i \pi
    \frac x a} \cos
  \sqrt{2}\phi_c \sin \sqrt{2} \phi_s,
\end{equation}
where we have assumed $ \langle n_\sigma\rangle =1/2$. 
Now, if we turn on the interchain hopping $\Omega$, we will
obtain with the help of (\ref{eq:boson-operator}) the following
bosonized form:
\begin{equation}
  \label{eq:transverse-field}
  H_{trans.} = \frac{2 \Omega}{\alpha}  \int dx \cos \sqrt{2} \theta_s [A_0^2  + 2 A_1^2 \cos
  \sqrt{8}\phi_c + 2 A_1^2 \cos \sqrt{8} \phi_s + \ldots],
\end{equation}
where $\ldots$ stands for less relevant operators. It is important to
note that in a renormalization group calculation, operators  $\cos \sqrt{8}\phi_c$ and $ \cos \sqrt{8} \phi_s$ are
generated in the flow. 
Since we are in the half-filled case, $\langle
n_\uparrow+n_\downarrow\rangle=1$, umklapp scattering $\cos \sqrt{8}
\phi_c$  is
present\cite{donohue_commensurate_bosonic_ladder,crepin2011} and  opens of a gap in the $c$ modes. 

For $\langle n_\uparrow \rangle
=\langle n_\downarrow \rangle$, the term $\cos \sqrt{8} \phi_\sigma$
is present, but is marginally irrelevant for repulsive interactions.

We have in the ground state $\langle \phi_c \rangle =0$. Using Eqs.~(\ref{eq:density-tot}), the
staggered and uniform components of the total density have exponentially decaying
correlations. Meanwhile, the
 antisymmetric density~(\ref{eq:spin-density})  has a simplified expression
for distances much larger than $u_c/\Delta_c$ where one can replace $\cos
  \sqrt{2}\phi_c$  by its average. \\
Besides density waves, the system can also present bond ordering.
The bond order wave order parameter $O_{BOW}^\sigma$
 for spin $\sigma$ boson is defined
by:
\begin{eqnarray}
  \label{eq:bow-def}
  b^\dagger_{j+1,\sigma} b_{j,\sigma} + b^\dagger_{j\sigma}
  b_{j+1,\sigma}= T(ja) + (-)^j O_{BOW}^\sigma(ja),
\end{eqnarray}
where $T$ is the kinetic energy density, and in bosonization
\begin{eqnarray}
  \label{eq:bow-spin}
  O_{BOW}^\sigma = C_0  \cos (2\phi_\sigma)
\end{eqnarray}
We can define the two order parameters:
\begin{eqnarray}\label{eq:bow-c}
  O_{BOW}^c &=& \sum_\sigma  O_{BOW}^\sigma = 2 C_0   \cos \sqrt{2}
  \phi_c \cos \sqrt{2}\phi_s \\
\label{eq:bow-s}
 O_{BOW}^s &=& \sum_\sigma \sign(\sigma) O_{BOW}^\sigma = 2 C_0   \sin \sqrt{2}
  \phi_c \sin \sqrt{2}\phi_s
\end{eqnarray}
In a Mott phase, $O_{BOW}^s$ is always short range ordered, while
$O_{BOW}^c \sim  \cos \sqrt{2}\phi_s$. If we consider the real space
Green's functions for the bosons, due to the long range order of
$\phi_c$, the exponentials $e^{i\beta\theta_c}$ of the dual field are
short range ordered, and the boson Green's functions decay
exponentially. Excitations of the total density are solitons and
antisolitons of
topological charge $\phi_c(+\infty)-\phi_c(-\infty)=\pm
\pi/\sqrt{2}$. Such an excitation corresponds to a change of particle
number $\pm 1$, and for fixed particle number density excitations are
formed of soliton/antisoliton pairs.

For $\lambda=0$, the term  $2 \Omega A_0^2 \cos \sqrt{2} \theta_s$ is a
relevant perturbation with scaling dimension $1/(2K_s)$, opening a
 gap $\Delta_s \sim \Omega^{2 K_s/(4K_s-1)}$ in the antisymmetric sector.
We have in the ground state  $\langle \theta_\sigma\rangle=\pi/\sqrt{2}$
and as a result all the fields $e^{i\beta \phi_\sigma}$ are short
range ordered.  Therefore, in the Mott-Meissner phase, all the
density wave and bond order wave order parameters are short range
ordered. Because of the long range order of $\theta_s$ we expect that
the correlation functions $\langle b_{j,\sigma} b^\dagger_{l,-\sigma}\rangle$ are
non-zero. A more
precise estimate of the gap (albeit without log corrections)
 can be obtained from the results of
\cite{zamolodchikov_energy_sg,lukyanov_sinegordon_correlations}.
Using these results, we predict that the soliton mass $\Delta_s$ behaves as:
\begin{eqnarray}
  \label{eq:gap-lukyanov}
  \Delta_s=\frac{u_s}{a} \frac{2 \Gamma\left(\frac {1}{8K_s-2}\right)}{\sqrt{\pi}
    \Gamma\left(\frac{2K_s}{4K_s-1}\right)} \left(\frac{\pi
      \Gamma\left(1-\frac{1}{4K_s}\right)}{\Gamma\left(\frac 1 {4K_s}\right)}
    \frac{\Omega A_0^2 a}{u_s}\right)^{\frac{2K_s}{4K_s-1}},
\end{eqnarray}
and the soliton/antisoliton dispersion is $E_s(k)=\sqrt{(u_sk)^2 +
  \Delta_s^2}$. In the case of hard core bosons, we have to set
$K_s=1$ in Eq.(\ref{eq:gap-lukyanov}). In such case,
besides the solitons and antisolitons, there are two
breathers\cite{uhrig_excitation_staggered,affleck86_dimerized,tsvelik_excitation_staggered},
a light breather of mass $\Delta_s$ and a heavy breather of mass
$\sqrt{3} \Delta_s$. The topological charge of the $\theta_s$ solitons/antisolitons
is $\theta_s(+\infty)-\theta_s(-\infty)=\pm \pi \sqrt{2}$. The
solitons therefore carry a spin current $u_s K_s$. In the case where one is considering
the gap between the ground state and an excited  state of total spin
current zero (i. e. containing at least one soliton and one
antisoliton), the measured gap will be $2\Delta_s$.
The amplitude $A_0$ can be estimated for hard core bosons in the case of
half-filling.\cite{ovchinnikov2004}
\begin{eqnarray}\label{eq:exact-amplitude-hcb}
  A_0^2= 2^{1/6} e^{1/2} A^{-6},
\end{eqnarray}
where $A\simeq 1.282427$ is Glaisher's constant.
Using (\ref{eq:gap-lukyanov}) with $K_s=1$ and
(\ref{eq:exact-amplitude-hcb}), we find:
 \begin{eqnarray}
  \label{eq:gap-half-filled}
  \Delta_s=\frac{u_s}{a} \frac{2 \Gamma(1/6)}{\sqrt{\pi}
    \Gamma(2/3)} \left(\frac{2^{1/6} e^{1/2} A^{-6} \Gamma(3/4) }{ \Gamma(1/4)}
    \frac{\pi \Omega a}{u_s}\right)^{2/3},
\end{eqnarray}
for half-filling. Using $u_s=2ta$, we finally have
$\Delta_s/t=3.3896(\Omega/t)^{2/3}$. The marginally irrelevant
operator $\cos \sqrt{8}\phi_s$ can give rise to logarithmic
corrections to that scaling\cite{oshikawa_cu_benzoate_short,oshikawa_cu_benzoate}
of the form $\Delta_s \sim  \Omega^{2/3}|\ln\Omega|^{1/6}$.

\subsection{Commensurate Incommensurate transition}
Neglecting the marginally irrelevant term $\cos \sqrt{8} \phi_s$ as in
\cite{orignac01_meissner},
 the Hamiltonian~(\ref{eq:bosonized-s})~(\ref{eq:transverse-field})  describes the
C-IC
transition\cite{japaridze_cic_transition,pokrovsky_talapov_prl,schulz_cic2d}.
When $\lambda$ exceeds the threshold $\lambda_c \sim (\Omega A_0^2
a/u_s)^{2-1/(2K_s)}$,
it becomes energetically favorable to populate the ground state
with a finite density of solitons of the field $\theta_s$ to form a
Tomonaga-Luttinger liquid (TLL)  of solitons.  The
low energy properties of that TLL are described by the effective Hamiltonian:
\begin{eqnarray}
  \label{eq:eff-cic}
  H^*= \int \frac{dx}{2\pi} \left[ u^*_s(\lambda) K^*_s(\lambda) (\pi \Pi^*_s)^2 +\frac {u^*_s(\lambda)} {K^*_s(\lambda)} (\partial_x \phi)^2 \right],
\end{eqnarray}
and we have $\pi \Pi_s(x) = \pi \Pi_s^*(x) - q(\lambda)/\sqrt{2}$;
$\theta_s(x)=\theta_s^*(x) - q(\lambda)x/\sqrt{2}$. Near
the transition point $\lambda_c$, we have $ q (\lambda) \sim C \sqrt{\lambda-\lambda_c}$. Moreover, as
$\lambda \to \lambda_c+0$, $K_s^*(\lambda)$ goes to a limiting value
$K_s^{(0)}$ such that\cite{schulz_cic2d,chitra_spinchains_field} the
scaling dimension of $\cos \sqrt{2} \theta_s$ becomes $1$. Since the
scaling dimension of $\cos \sqrt{2} \theta_s$ with a Hamiltonian of
the form~(\ref{eq:eff-cic}) is $1/[2K_s^*(\lambda)]$ we have
$K_s^{*}(\lambda \to \lambda_c+0)=1/2$. Using a fermionization
method\cite{orignac01_meissner},
an explicit form of $q(\lambda)$ can be obtained for $K_s=1/2$.

The antisymmetric leg current (or screening current) operator  
\begin{eqnarray}
J_s(j)=-it\sum_{\sigma}\left(\sigma
e^{i\lambda\sigma}  b^\dagger_{j,\sigma}b_{j+1,\sigma} -\sigma
e^{-i\lambda\sigma} b^\dagger_{j+1,\sigma}b_{j,\sigma} 
\right), 
\label{eq:Js}
\end{eqnarray}

 is obtained by differentiating the Hamiltonian with
respect to the parameter $\lambda$. One finds:
\begin{eqnarray}
  J_s(x)=\frac{u_s K_s}{\pi \sqrt{2}} \left(\pi \Pi_s +
    \frac{\lambda} {a\sqrt{2}}  \right)
\end{eqnarray}
In the Meissner phase, $\langle  \Pi_s \rangle = 0$ so that:
\begin{eqnarray}
 \langle  J_s \rangle =\frac {u_s K_s}{2 \pi} \lambda
\end{eqnarray}
In  the vortex phase, $J_s = u_s K_s (\lambda -\sign(\lambda)
q(\lambda))/(2\pi)$. By fermionization\cite{orignac01_meissner},
$q(\lambda)=\sqrt{\lambda^2-\lambda_c^2}$ is obtained for $K_s=1/2$.

In the Mott-Vortex state, we obtain the rung current as:
\begin{eqnarray}
  \label{eq:conversion}
  j_\perp(x)=\Omega A_0^2 \sin [\sqrt{2} \theta_s^*(x) -q(\lambda)x].
\end{eqnarray}
In both  the Meissner and the vortex phase, its expectation value
$\langle j_\perp\rangle =0$  vanishes. In the Meissner phase, the
conversion current correlation function $C(x)=\langle  j_\perp (x)  j_\perp (0)\rangle$ decays exponentially
whereas in the vortex phase:
\begin{eqnarray}
  C(x)= \frac 1 2 (\Omega A_0^2)^2  \left(\frac{a^2}{x^2 + a^2}\right)^{\frac 1
    {2K_s^*}} \cos [q(\lambda) x]
\end{eqnarray}
For $K_s^*=1/2$, the Fourier transform:
\begin{equation}
  C(k)= \frac{a  (\Omega A_0^2)^2}{4} (e^{-|k-q(\lambda)|a} +
  e^{-|k+q(\lambda)|a}),
\end{equation}
has two cusps at $k=\pm q(\lambda)$. Peaks divergent with system size
appear in $C(k)$ for $K_s^*>1$.

We can also obtain the spin-spin and bond order correlation functions.
In the Mott-Vortex phase, the fields $e^{i\beta \phi_\sigma}$ have
quasi-long range order. As a result, we find that:
\begin{eqnarray}
  \label{eq:mott-meissner}
  \langle O_{BOW}^c(x) O_{BOW}^c(0)\rangle &=& C_0^2 (\langle \cos
  \sqrt{2} \phi_c \rangle)^2 \left(\frac
    {a^2}{x^2+a^2}\right)^{K_s^*/2} \\
\langle \sigma^z(x) \sigma^z(0)\rangle &=& \frac{K_s}{4\pi^2}
\frac{a^2-x^2}{(x^2+a^2)^2} + (-)^{\frac x a} B_0^2  (\langle \cos
  \sqrt{2} \phi_c \rangle)^2 \left(\frac
    {a^2}{x^2+a^2}\right)^{K_s^*/2}
\end{eqnarray}
If we turn to the Fourier transforms,%
we find that for $k\simeq 0$,
\begin{eqnarray}
  \label{eq:S_spin-small-k}
  S(k)=\frac{K_s^*|k|}{4\pi} e^{-|k|a},
\end{eqnarray}
by using the integral $\int_{-\infty}^\infty  \frac {dx e^{ikx}} {x^2+a^2} = \frac \pi a
  e^{-|k| a}$. For $k \simeq \frac \pi a$ and $K_s^*<1$,  we find that the correlation functions
$S(k)$ and $S_{BOWc}(k)$ are divergent  as:
\begin{eqnarray}\label{eq:exponent-S}
  S (k) \sim  S_{bow}^c (k)\sim \left|k -\frac \pi a \right|^{K_s^*-1},
\end{eqnarray}
and as a result a divergence going as $|ka|^{-1/2}$ is expected at the
transition, while far from the transition $K_s^*\simeq 1$, giving only
a weak power law or logarithmic divergence. 
For $K_s^*>1$, both $ S_s(k)$ and for $S_{bow}^c(k)$ remain finite in
the vicinity of $\pi/a$.
\begin{figure}[h]
  \centering
  \includegraphics[width=9.5cm]{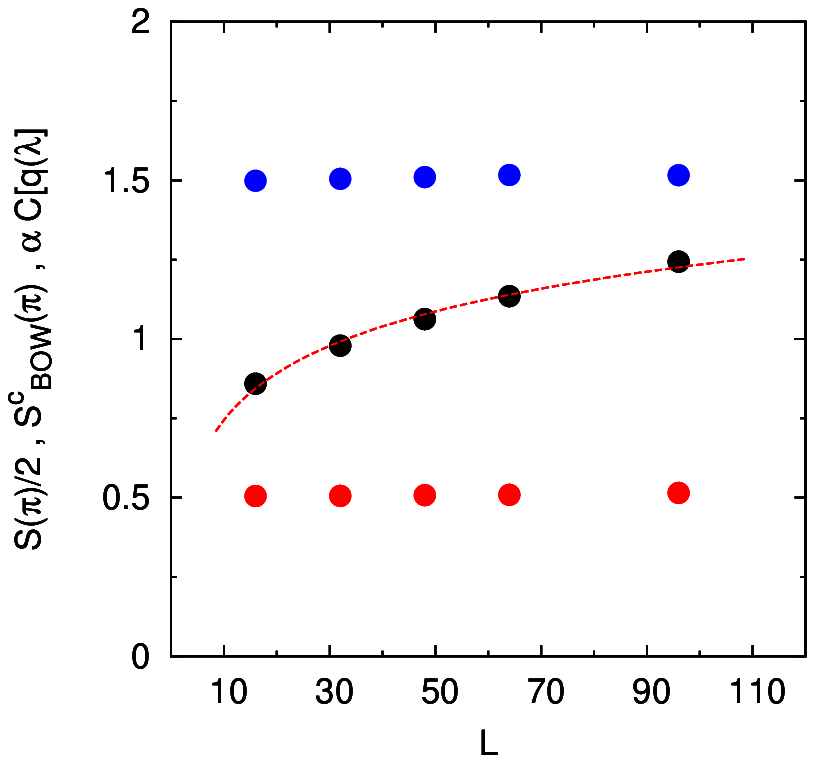}
  \caption{Size dependence of  $S(k=\pi)$ (red dots) and
    $S^c_{BOW}(k=\pi)$ (blue dots), for  system sizes $L=16,32,48,64$
    and $96$.  Data are for
    $\Omega=0.25$ and $\lambda=\pi/2$, well inside the vortex phase
    region. Both quantities don't show a visible size dependence in
    agreement with (\ref{eq:exponent-S}) for $K_s^*=1$.  Solid black dots are $C(k=q(\lambda))\alpha$ with $\alpha=\Omega^2/16$
    to be on the same scale on the graph. The size-scaling is
    compatible with a logarithmic dependence (red dotted line) as expected far from the transition region} 
  \label{fig:scaling}
\end{figure}

If we turn to the momentum distribution, we will find:
\begin{eqnarray}
  \langle b_{j,\sigma} b^\dagger_{l,\sigma'} \rangle &=& \delta_{\sigma
    \sigma'} e^{-i \sigma q(\lambda)a (j-l) }
  \langle e^{i \theta_c(ja)/\sqrt{2}}  e^{-i
    \theta_c(la)/\sqrt{2}}\rangle \langle \langle e^{i \theta_s^*(ja)/\sqrt{2}}  e^{-i
    \theta_s(la)/\sqrt{2}}\rangle \\
  &=& \delta_{\sigma
    \sigma'} e^{-i \sigma q(\lambda)a (j-l)}\frac{\langle e^{i \theta_c(ja)/\sqrt{2}}  e^{-i
    \theta_c(la)/\sqrt{2}}\rangle}{\left(1+(x/a)^2\right)^{1/(8K_s^*)}}.
\end{eqnarray}
because of the exponential decay of the charge correlator, we expect
that the momentum distribution of spin $\sigma$ particles will be
centered around $k= -\sigma q(\lambda)$. The total momentum
distribution will thus have two peaks for $k=\pm q(\lambda)/2$. 

\section{Incommensuration for $\lambda \simeq \pi$}
\label{sec:lambda-pi}

For $\lambda$ close to $\pi$, the form (\ref{eq:bosonized-s}) for the
Hamiltonian cannot be used as $\lambda u_s/a$ is not a small quantity
compared with the energy cutoff $u_s/a$. To describe the low energy
physics at $\lambda=\pi$,
it is necessary to choose a gauge with the vector potential
along the rungs of the ladder, so that the interchain hopping reads:
\begin{equation}
  \label{eq:gauge-rung}
  H_{hop.}=\Omega \sum_j (-)^j b^\dagger_{j,\sigma} b_{j,-\sigma},
\end{equation}

\begin{figure}[h]
  \centering
  \includegraphics[width=9.5cm]{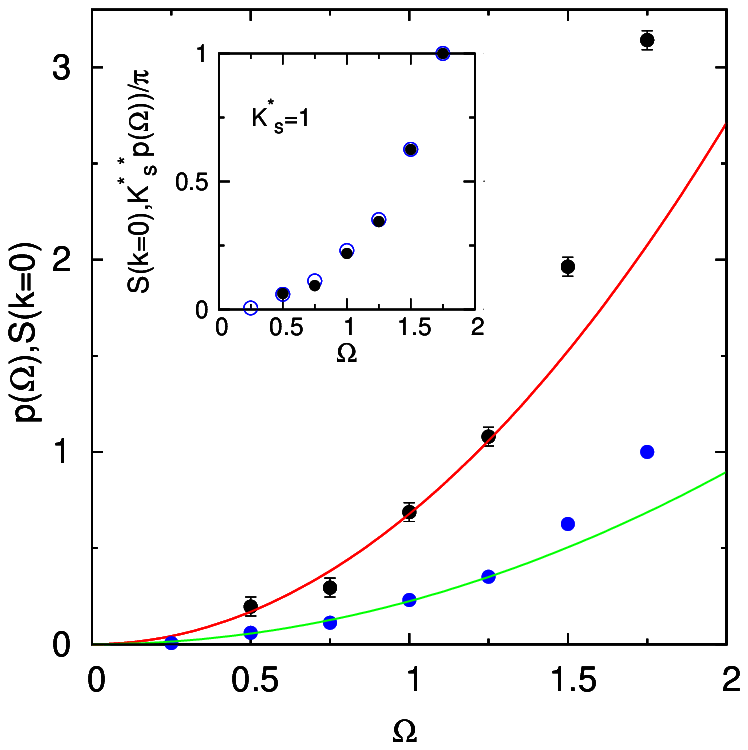}
  \caption{A graph of the incommensuration $p(\Omega)=h_s/u_s$ at
    $\lambda=\pi$ obtained from  numerical
    data (L=64 in PBC). The blue dots are the value of $S(k=0)$, while the
    black dots correspond to the slope discontinuity
    $k=p(\Omega)$. The red and green lines are quadratic fits. In the inset
    we show $S(k=0)$, open blue dots, together with 
    the $K^*_s p(\Omega)/\pi$, black soid dots,
    with the choice $K^*_s=1$ }
  \label{fig:qomega}
\end{figure}

Applying bosonization to (\ref{eq:gauge-rung}), we obtain the
following representation for interchain hopping:
\begin{eqnarray}
  \label{eq:rung-pi}
  H_{hop.}=\frac{\Omega}{2\pi a} \int dx \cos \sqrt{2} \phi_c \left[
    e^{-i\sqrt{2}(\theta_s +\phi_s)} +e^{-i\sqrt{2}(\theta_s -\phi_s)}
+ e^{i\sqrt{2}(\theta_s +\phi_s)} +e^{i\sqrt{2}(\theta_s
  -\phi_s)}\right],
\end{eqnarray}
which can be rewritten in terms of $\mathrm{SU(2)}_1$
Wess-Zumino-Novikov-Witten (WZNW) currents\cite{tsvelik_book}:
\begin{eqnarray}
  \label{eq:rung-pi-wzw}
  H_{hop.}=\Omega \int dx \cos \sqrt{2} \phi_c (J_R^y + J_L^y).
\end{eqnarray}
The resulting Hamiltonian $H_c+H_s+H_{hop.}$
 can be treated in mean-field
 theory\cite{nersesyan_incom,lecheminant2001,jolicoeur2002,zarea04_chiral_xxz},
 giving:
\begin{eqnarray}\label{eq:rung-pi-mf}
H_{MF}&=&H_c^{MF}+H_s^{MF}, \\
\label{eq:rung-pi-mf-c}
H_c^{MF}&=& \int \frac{dx}{2\pi} \left[ u_c K_c (\pi \Pi_c)^2 +\frac
    {u_c} {K_c} (\partial_x \phi_c)^2 \right] + \frac{g_c}{\pi a} \int
  dx \cos \sqrt{2} \phi_c, \\
\label{eq:rung-pi-mf-s}
H_s^{MF}&=& \frac {2\pi u_s}{3} \int dx (\mathbf{J}_R\cdot
\mathbf{J}_R+ \mathbf{J}_L\cdot \mathbf{J}_L) + h_s \int dx (J_R^y + J_L^y),
\end{eqnarray}
where:
\begin{eqnarray}
  \label{eq:rung-pi-selfcon}
  \frac{g_c}{\pi a} &=& 8 \Omega \langle J_R^y + J_L^y \rangle_{s,MF},\nonumber \\
  h_s &=&  8 \Omega \langle \cos \sqrt{2} \phi_c\rangle_{c,MF}.
\end{eqnarray}
Using a $\frac \pi 2$ rotation around the $x$ axis, $J_{\nu}^y
=\tilde{J}_{\nu}^z$, $J_\nu^z=- \tilde{J}_{\nu}^y$, and applying
abelian bosonization\cite{nersesyan_2ch}, we rewrite:
\begin{eqnarray}
  \label{eq:rung-pi-mf-rotated}
  H_s^{MF}&=&\int \frac{dx}{2\pi} u_s \left[(\pi \tilde{\Pi}_s)^2 +
    (\partial \tilde{\phi}_s)^2 \right] - \frac {h_s}{\pi \sqrt{2}}
  \int \partial_x \tilde{\phi}_s dx ,
\end{eqnarray}
which allows us to write:
\begin{eqnarray}
  - \frac {1}{\pi \sqrt{2}} \langle \partial_x \tilde{\phi}_s\rangle =
  \sum_{\nu=R,L} \langle \tilde{J}_{\nu}^z \rangle =  \langle J_R^y +
  J_L^y  \rangle = -\frac {h_s}{2\pi u_s},
\end{eqnarray}
and allows us to solve (\ref{eq:rung-pi-selfcon}) with $h_s \sim
\Omega^2$ and $g_c \sim \Omega^3$. We obtain a gap in the total
density excitations, $\Delta_c \sim \Omega^2$, while the antisymmetric
modes remain gapless and develop an incommensuration. To characterize
the incommensuration, we need to detail the rotation of the
$\mathrm{SU(2)}_1$ WZNW currents and primary fields. After shifting
$\tilde{\phi}_s \to  \tilde{\phi}_s + \frac{h_s x}{u_s \sqrt{2}}$,
we find:
\begin{eqnarray}
  \label{eq:rotation-currents}
  -\frac 1 {\pi \sqrt{2}} \partial_x \phi_s = -\frac 1 {2 \pi a}
  \sum_{r,r'=\pm} e^{i r\sqrt{2} (\tilde{\theta}_s+  r'\tilde{\phi}_s) + i rr'
      \frac{h_s x}{u_s}}, \\
  \sum_{r=\pm} \sin \sqrt{2} (\theta_s + r \phi_s) = \sum_{r=\pm} \sin
  \sqrt{2} \left(\tilde{\theta}_s + r \tilde{\phi}_s + r \frac{h_s
      x}{u_s} \right), \\
 \sum_{r=\pm} \cos \sqrt{2} (\theta_s + r \phi_s) =  -\frac 1 {\pi
   \sqrt{2}} \partial_x \tilde{\phi}_s -\frac {h_s}{2\pi u_s},
\end{eqnarray}
and:
\begin{eqnarray}
  \label{eq:rotation-primaries}
  \sin \sqrt{2} \theta_s &=& \sin \sqrt{2} \tilde{\theta}_s \\
  \cos \sqrt{2} \theta_s &=& \sin \left(\sqrt{2} \tilde{\phi}_s +
    \frac{h_s x}{u_s}\right) \\
  \sin \sqrt{2} \phi_s &=& - \cos \sqrt{2} \tilde{\theta}_s \\
  \cos \sqrt{2} \phi_s &=& \cos \left(\sqrt{2} \tilde{\phi}_s +
    \frac{h_s x}{u_s}\right)
\end{eqnarray}
Since we have:
\begin{eqnarray}
  j_\perp(j) &=&  \frac \Omega {\pi a} \sum_{r=\pm} \sin \sqrt{2}
  (\theta_s + r \phi_s) +  \frac{2\Omega (-)^j }{\pi a} \sin \sqrt{2} \theta_s, \\
 \sigma^z(x) &=&-\frac 1 {\pi \sqrt{2}} \partial_x \phi_s + \frac{(-1)^j}{\pi a} \langle \cos \sqrt{2} \phi_c
\rangle \sin \sqrt{2} \phi_s, \\
O_{BOW}^c &=& \frac{(-1)^j}{\pi a} \langle \cos \sqrt{2} \phi_c
\rangle \cos \sqrt{2} \phi_s
\end{eqnarray}
we find, after the rotation:
\begin{eqnarray}
  j_\perp(j) = \frac{\Omega}{\pi a} \sum_{r=\pm} \sin
  \sqrt{2} \left(\tilde{\theta}_s + r \tilde{\phi}_s + r \frac{h_s
      x}{u_s} \right) + \frac{2\Omega (-)^j }{\pi a} \sin \sqrt{2} \tilde{\theta}_s, \\
 \sigma^z(x) = \frac{1}{\pi a} \sum_{r=\pm} \cos
  \sqrt{2} \left(\tilde{\theta}_s + r \tilde{\phi}_s + r \frac{h_s
      x}{u_s} \right) - \frac{(-1)^j}{\pi a} \langle \cos \sqrt{2} \phi_c
\rangle \cos \sqrt{2} \tilde{\theta}_s, \\
 O_{BOW}^c = \frac{(-1)^j}{\pi a} \langle \cos \sqrt{2} \phi_c
\rangle \cos \left(\sqrt{2} \tilde{\phi}_s +
    \frac{h_s x}{u_s}\right)
\end{eqnarray}
so that:
\begin{eqnarray}
  \langle  j_\perp(j)  j_\perp(j')\rangle &\sim& \frac 1 {2 \pi^2
    (j-j')^2} \cos \left(\frac{h_s (j-j')}{u_s} \right) +
  \frac{(-1)^{j-j'}}{|j-j'|}, \\
  \langle \sigma^z(j) \sigma^z (j') \rangle  &\sim&\frac 1 {2 \pi^2
    (j-j')^2} \cos \left(\frac{h_s (j-j')}{u_s} \right) +
  \frac{(-1)^{j-j'}}{|j-j'|}, \\
    \langle O^c_{BOW}(j) O^c_{BOW} (j') \rangle  &\sim&
    \frac{(-1)^{j-j'}}{|j-j'|}  \cos \left(\frac{h_s (j-j')}{u_s}
    \right)
\end{eqnarray}
We see that an incommensuration of wavevector $p(\Omega)=h_s/u_s$ develops in the 
$k\simeq 0$ component of the rung current and density wave correlations. Since $p(\Omega) \sim
\Omega^2$ (see Fig.~\ref{fig:qomega}), the incommensuration increases with interchain hopping.
The Fourier transform of the $k\simeq 0$ component behaves as
$|k-p(\Omega)|+|k+p(\Omega)|$, i. e. it is constant for $|k|<p(\Omega)$
and linear in $k$ for $|k|>p(\Omega)$.

Since the boson annihilation operators do not correspond to primary
fields of the $SU(2)_1$ WZNW model, we cannot directly derive their
expression from $SU(2)$ symmetry. However, since
$e^{i\theta_s/\sqrt{2}}$ has conformal  dimensions $(1/16,1/16)$ its expression in
  terms of $\tilde{\theta}_s$ and $\tilde{\phi}_s$ has to be a sum of
 operators of conformal dimensions $(1/16,1/16)$. A general expression is:
 \begin{eqnarray}
  e^{i\frac{\theta_s}{\sqrt{2}}} = A_0
  e^{i\frac{\tilde{\theta}_s}{\sqrt{2}}} + A_1
  e^{i\frac{\tilde{\phi}_s}{\sqrt{2}}} + A_2
  e^{-i\frac{\tilde{\theta}_s}{\sqrt{2}}}+  A_3
  e^{-i\frac{\tilde{\phi}_s}{\sqrt{2}}}.
 \end{eqnarray}
Moreover, the operator product expansion:
\begin{eqnarray}
  e^{i\frac{\theta_s(x)}{\sqrt{2}}}
  e^{i\frac{\theta_s(x')}{\sqrt{2}}} = \left|\frac{x-x'}
    a\right|^{1/4} e^{i {\sqrt{2} \theta_s(x)}},
\end{eqnarray}
has to be satisfied, so, for $\lambda=\pi$,  we must have $\{A_k,A_j\}=\delta_{kj} e^{-i k
  \frac \pi 2} $ which can be satisfied by writing $A_k$ as the
product of a $4\times 4$ Dirac matrix\cite{itzykson-zuber} by a phase
$e^{-ik \frac \pi 4}$. In the case of $\lambda=\pi$ we obtain the
correlator:
\begin{eqnarray}
 \langle e^{i\frac{\theta_s(x)}{\sqrt{2}}}
  e^{-i\frac{\theta_s(x')}{\sqrt{2}}} \rangle = \left(\frac a
    {|x-x'|}\right)^{\frac 1 4} \left[2 + 2 \cos \frac {h_s}{u_s}
    (x-x') \right],
\end{eqnarray}
where the $\cos$ results from the correlation of the $e^{\pm i
  \tilde{\phi}_s/\sqrt{2}}$. This implies that the momentum
  distribution is:
  \begin{eqnarray}
    n_\sigma(k)&=&\int dx e^{i(k+\pi \sigma)x} \langle e^{-i
        \frac{\theta_c}{\sqrt{2}}(x)}   e^{i
        \frac{\theta_c}{\sqrt{2}}(0)} \rangle  \langle e^{i\frac{\theta_s(x)}{\sqrt{2}}}
  e^{-i\frac{\theta_s(0)}{\sqrt{2}}} \rangle \\
      &=& \int dx e^{i(k+\pi \sigma)x}  \langle e^{-i
        \frac{\theta_c}{\sqrt{2}}(x)}   e^{i
        \frac{\theta_c}{\sqrt{2}}(0)} \rangle  \left(\frac a
    {|x|}\right)^{\frac 1 4} \left[2 + 2 \cos \frac {h_s x}{u_s}
     \right] \\
     &=& \nu(k+\pi \sigma)+ \frac 1 2  \nu\left(k+\pi \sigma +  \frac
     {h_s}{u_s}\right)     + \frac 1 2 \nu\left(k+\pi \sigma -  \frac
     {h_s}{u_s}\right), 
  \end{eqnarray}
where:
\begin{equation}
  \label{eq:nu-def}
  \nu(k)=\int dx e^{ik x}  \langle e^{-i
        \frac{\theta_c}{\sqrt{2}}(x)}   e^{i
        \frac{\theta_c}{\sqrt{2}}(0)} \rangle  \left(\frac a
    {|x|}\right)^{\frac 1 4}. 
\end{equation}
Since the width at half maximum of the Lorentzian-shaped graph of the function $\nu$
scales as $\Delta_c/u_c \sim \Omega^2$ and $h_s/u_s\sim \Omega^2$, the
graph of  $n(k)$ can either comprise 3 peaks or a single broad peak
centered in $\pi \sigma$ depending on the dimensionless ratio $h_s
u_c/(u_s \Delta_c)$.

When $\lambda <\pi$, we choose a gauge such that:
\begin{eqnarray}
  H&=&-t \sum_{j,\sigma} \left( b^\dagger_{j,\sigma}
  e^{i(\lambda-\pi)\sigma}b_{j+1,\sigma} + \mathrm{H.c.} \right)
\nonumber \\
 && + \Omega \sum_{j,\sigma} (-1)^j b^\dagger_{j,\sigma} b_{j,-\sigma},
\end{eqnarray}
and we define $\delta \lambda = \lambda -\pi$. The mean field
Hamiltonian becomes $H_{MF}=H_c^{MF}+H_s^{MF}$ with:
\begin{eqnarray}
  \label{eq:hs-mf-pi-lambda}
  H_s^{MF}= \frac {2\pi u_s}{3} \int dx (\mathbf{J}_R\cdot
\mathbf{J}_R+ \mathbf{J}_L\cdot \mathbf{J}_L) + h_s \int dx (J_R^y +
J_L^y) +\frac{u_s \delta\lambda}{a} \int dx (J_R^z -J_L^z),
\end{eqnarray}
and $H_c^{MF}$ unchanged. We now have to make a different rotation
around $x$ for
the right moving and the left moving current:
\begin{eqnarray}
  \label{eq:chiral-rotation}
  J_R^y &=& \sin \varphi \tilde{J}_R^y + \cos \varphi  \tilde{J}_R^z
  \\
 J_R^z &=& -\cos \varphi \tilde{J}_R^y + \sin \varphi  \tilde{J}_R^z
  \\
  J_L^y &=& - \sin \varphi \tilde{J}_L^y + \cos \varphi  \tilde{J}_L^z
  \\
 J_L^z &=& -\cos \varphi \tilde{J}_L^y + - \sin \varphi  \tilde{J}_L^z
\end{eqnarray}
To find;
\begin{eqnarray}
   H_s^{MF}&=&\int \frac{dx}{2\pi} u_s \left[(\pi \tilde{\Pi}_s)^2 +
    (\partial \tilde{\phi}_s)^2 \right] - \frac {h_s(\lambda)}{\pi \sqrt{2}}
  \int \partial_x \tilde{\phi}_s dx ,
\end{eqnarray}
where $h_s(\lambda)=\sqrt{h_s^2 + u_s^2 (\delta \lambda/a)^2 }$. We still have $\langle J_R^y + J_L^y \rangle = -\frac{h_s}{2\pi u_s}$, so the mean-field equations remain the same.
We also find:
\begin{eqnarray}
  \label{eq:chiral-primaries}
   \sin \sqrt{2} \theta_s &=& \frac{h_s}{h_s(\lambda)} \sin \sqrt{2} \tilde{\theta}_s +\frac{u\delta
       \lambda/a}{h_s(\lambda)} \cos \left(\sqrt{2} \tilde{\phi}_s + \frac{h_s(\lambda)}{u_s} x \right)  \\
  \cos \sqrt{2} \theta_s &=& \sin \left(\sqrt{2} \tilde{\phi}_s + \frac{h_s(\lambda)}{u_s}  x\right) \\
  \sin \sqrt{2} \phi_s &=& - \cos \sqrt{2} \tilde{\theta}_s \\
  \cos \sqrt{2} \phi_s &=& \frac{h_s}{h_s(\lambda)} \cos \left(\sqrt{2} \tilde{\phi}_s +
    \frac{h_s(\lambda)}{u_s} x\right) -  \frac{u\delta
       \lambda/a}{h_s(\lambda)}  \sin \sqrt{2} \tilde{\theta}_s
\end{eqnarray}

The staggered part of the rung current correlations becomes:
\begin{eqnarray}
  \langle j_\perp(x) j_\perp(0) \rangle \sim \frac{(-)^{x/a}}{|x|}
  \frac{h_s^2+ \left(\frac{u_s\delta\lambda}{a}\right)^2 \cos \frac{h_s(\lambda)}{u_s} x}{h_s(\lambda)^2},
\end{eqnarray}
so that the Fourier transform will present peaks at $k=\frac \pi a$
and $k=\frac \pi a 
\pm h_s(\lambda)/u_s$. When $\delta \lambda$
is increased, the two peaks at  $k=\frac \pi a 
\pm \frac {h_s(\lambda)}{u_s}$ become dominant, and we
crossover to the behavior already discussed for weak $\lambda$.
In the case of $S(k)$, the peak at $k=\pi$  is not split as
$\lambda$ is reduced. If we look at the $BOW^c$ correlations, a peak
at $k=\pi$ appears, and becomes the dominant peak when $\delta
\lambda$ is increased.
Using the rotation (\ref{eq:chiral-rotation}) we can also obtain the
antisymmetric density correlations as:
\begin{eqnarray}
  \frac 1 {2\pi^2} \langle \partial_x \phi_s(x) \partial_x \phi_s(0)
  \rangle = \frac 1 {2\pi^2 x^2} \left[ \frac{\left(\frac{u_s\delta\lambda}{a}\right)^2}{h_s^2(\lambda)} + \frac{ h_s^2 }{h_s^2 +
    \left(\frac{u_s\delta\lambda}{a}\right)^2} \cos
  \left(h_s(\lambda)/u_s x\right)\right]
\end{eqnarray}
When $\delta \lambda$ increases, this expression crosses over to the
$1/(2\pi^2 x^2)$ which was obtained at small $\lambda$. $S(k)$ now
presents a change of slope at $|k|=h_s(\lambda)/u_s$.  

As to the $k\simeq 0$ component of the rung current, since it is
proportional to $J_R^x+J_L^x$ it becomes $\tilde{J_R}^x+\tilde{J}_L^x$
under the rotation, and the correlator becomes:
\begin{eqnarray}
  \frac 1 {2\pi^2 (j-j')^2} \cos \left ( \frac{h_s(\lambda)}{u_s}  (j-j') \right),
\end{eqnarray}


\begin{thebibliography}{44}
\expandafter\ifx\csname natexlab\endcsname\relax\def\natexlab#1{#1}\fi
\expandafter\ifx\csname bibnamefont\endcsname\relax
  \def\bibnamefont#1{#1}\fi
\expandafter\ifx\csname bibfnamefont\endcsname\relax
  \def\bibfnamefont#1{#1}\fi
\expandafter\ifx\csname citenamefont\endcsname\relax
  \def\citenamefont#1{#1}\fi
\expandafter\ifx\csname url\endcsname\relax
  \def\url#1{\texttt{#1}}\fi
\expandafter\ifx\csname urlprefix\endcsname\relax\def\urlprefix{URL }\fi
\providecommand{\bibinfo}[2]{#2}
\providecommand{\eprint}[2][]{\url{#2}}

\bibitem[{\citenamefont{Tinkham}(1975)}]{tinkham_book_superconductors}
\bibinfo{author}{\bibfnamefont{M.}~\bibnamefont{Tinkham}},
  \emph{\bibinfo{title}{Introduction to Superconductivity}}
  (\bibinfo{publisher}{McGraw Hill}, \bibinfo{address}{New York},
  \bibinfo{year}{1975}).

\bibitem[{\citenamefont{Mermin and Wagner}(1967)}]{mermin_wagner_theorem}
\bibinfo{author}{\bibfnamefont{N.~D.} \bibnamefont{Mermin}} \bibnamefont{and}
  \bibinfo{author}{\bibfnamefont{H.}~\bibnamefont{Wagner}},
  \bibinfo{journal}{Phys. Rev. Lett.} \textbf{\bibinfo{volume}{17}},
  \bibinfo{pages}{1133} (\bibinfo{year}{1967}).

\bibitem[{\citenamefont{Hohenberg}(1967)}]{hohenberg67_theorem}
\bibinfo{author}{\bibfnamefont{P.~C.} \bibnamefont{Hohenberg}},
  \bibinfo{journal}{Phys. Rev.} \textbf{\bibinfo{volume}{158}},
  \bibinfo{pages}{383} (\bibinfo{year}{1967}).

\bibitem[{\citenamefont{Kardar}(1986)}]{kardar_josephson_ladder}
\bibinfo{author}{\bibfnamefont{M.}~\bibnamefont{Kardar}},
  \bibinfo{journal}{Phys. Rev. B} \textbf{\bibinfo{volume}{33}},
  \bibinfo{pages}{3125} (\bibinfo{year}{1986}).

\bibitem[{\citenamefont{Orignac and Giamarchi}(2001)}]{orignac01_meissner}
\bibinfo{author}{\bibfnamefont{E.}~\bibnamefont{Orignac}} \bibnamefont{and}
  \bibinfo{author}{\bibfnamefont{T.}~\bibnamefont{Giamarchi}},
  \bibinfo{journal}{Phys. Rev. B} \textbf{\bibinfo{volume}{64}},
  \bibinfo{pages}{144515} (\bibinfo{year}{2001}), \eprint{cond-mat/0011497}.

\bibitem[{\citenamefont{Cha and Shin}(2011)}]{cha2011}
\bibinfo{author}{\bibfnamefont{M.-C.} \bibnamefont{Cha}} \bibnamefont{and}
  \bibinfo{author}{\bibfnamefont{J.-G.} \bibnamefont{Shin}},
  \bibinfo{journal}{Phys. Rev. A} \textbf{\bibinfo{volume}{83}},
  \bibinfo{pages}{055602} (\bibinfo{year}{2011}).

\bibitem[{\citenamefont{Japaridze and
  Nersesyan}(1978)}]{japaridze_cic_transition}
\bibinfo{author}{\bibfnamefont{G.~I.} \bibnamefont{Japaridze}}
  \bibnamefont{and} \bibinfo{author}{\bibfnamefont{A.~A.}
  \bibnamefont{Nersesyan}}, \bibinfo{journal}{JETP Lett.}
  \textbf{\bibinfo{volume}{27}}, \bibinfo{pages}{334} (\bibinfo{year}{1978}).

\bibitem[{\citenamefont{Pokrovsky and Talapov}(1979)}]{pokrovsky_talapov_prl}
\bibinfo{author}{\bibfnamefont{V.~L.} \bibnamefont{Pokrovsky}}
  \bibnamefont{and} \bibinfo{author}{\bibfnamefont{A.~L.}
  \bibnamefont{Talapov}}, \bibinfo{journal}{Phys. Rev. Lett.}
  \textbf{\bibinfo{volume}{42}}, \bibinfo{pages}{65} (\bibinfo{year}{1979}).

\bibitem[{\citenamefont{Granato}(1990)}]{granato_josephson_ladder}
\bibinfo{author}{\bibfnamefont{E.}~\bibnamefont{Granato}},
  \bibinfo{journal}{Phys. Rev. B} \textbf{\bibinfo{volume}{42}},
  \bibinfo{pages}{4797} (\bibinfo{year}{1990}).

\bibitem[{\citenamefont{Nishiyama}(2000)}]{nishiyama_josephson_ladder}
\bibinfo{author}{\bibfnamefont{Y.}~\bibnamefont{Nishiyama}},
  \bibinfo{journal}{Eur. Phys. J. B} \textbf{\bibinfo{volume}{17}},
  \bibinfo{pages}{295} (\bibinfo{year}{2000}), \eprint{cond-mat/0006311}.

\bibitem[{\citenamefont{{Dhar} et~al.}(2012)\citenamefont{{Dhar}, {Maji},
  {Mishra}, {Pai}, {Mukerjee}, and {Paramekanti}}}]{dhar2012}
\bibinfo{author}{\bibfnamefont{A.}~\bibnamefont{{Dhar}}},
  \bibinfo{author}{\bibfnamefont{M.}~\bibnamefont{{Maji}}},
  \bibinfo{author}{\bibfnamefont{T.}~\bibnamefont{{Mishra}}},
  \bibinfo{author}{\bibfnamefont{R.~V.} \bibnamefont{{Pai}}},
  \bibinfo{author}{\bibfnamefont{S.}~\bibnamefont{{Mukerjee}}},
  \bibnamefont{and}
  \bibinfo{author}{\bibfnamefont{A.}~\bibnamefont{{Paramekanti}}},
  \bibinfo{journal}{Phys. Rev. A} \textbf{\bibinfo{volume}{85}},
  \bibinfo{pages}{041602} (\bibinfo{year}{2012}).

\bibitem[{\citenamefont{{Dhar} et~al.}(2013)\citenamefont{{Dhar}, {Mishra},
  {Maji}, {Pai}, {Mukerjee}, and {Paramekanti}}}]{dhar2013}
\bibinfo{author}{\bibfnamefont{A.}~\bibnamefont{{Dhar}}},
  \bibinfo{author}{\bibfnamefont{T.}~\bibnamefont{{Mishra}}},
  \bibinfo{author}{\bibfnamefont{M.}~\bibnamefont{{Maji}}},
  \bibinfo{author}{\bibfnamefont{R.~V.} \bibnamefont{{Pai}}},
  \bibinfo{author}{\bibfnamefont{S.}~\bibnamefont{{Mukerjee}}},
  \bibnamefont{and}
  \bibinfo{author}{\bibfnamefont{A.}~\bibnamefont{{Paramekanti}}},
  \bibinfo{journal}{Phys. Rev. B} \textbf{\bibinfo{volume}{87}},
  \bibinfo{pages}{174501} (\bibinfo{year}{2013}).

\bibitem[{\citenamefont{{Petrescu} and {Le Hur}}(2013)}]{petrescu2013}
\bibinfo{author}{\bibfnamefont{A.}~\bibnamefont{{Petrescu}}} \bibnamefont{and}
  \bibinfo{author}{\bibfnamefont{K.}~\bibnamefont{{Le Hur}}},
  \bibinfo{journal}{Phys. Rev. Lett.} \textbf{\bibinfo{volume}{111}},
  \bibinfo{pages}{150601} (\bibinfo{year}{2013}).

\bibitem[{\citenamefont{Tokuno and Georges}(2014)}]{tokuno2014}
\bibinfo{author}{\bibfnamefont{A.}~\bibnamefont{Tokuno}} \bibnamefont{and}
  \bibinfo{author}{\bibfnamefont{A.}~\bibnamefont{Georges}},
  \bibinfo{journal}{New J. Phys.} \textbf{\bibinfo{volume}{16}},
  \bibinfo{pages}{073005} (\bibinfo{year}{2014}).

\bibitem[{\citenamefont{Petrescu and Le~Hur}(2015)}]{petrescu2015}
\bibinfo{author}{\bibfnamefont{A.}~\bibnamefont{Petrescu}} \bibnamefont{and}
  \bibinfo{author}{\bibfnamefont{K.}~\bibnamefont{Le~Hur}},
  \bibinfo{journal}{Phys. Rev. B} \textbf{\bibinfo{volume}{91}},
  \bibinfo{pages}{054520} (\bibinfo{year}{2015}).

\bibitem[{\citenamefont{Kele\c{s} and Oktel}(2015)}]{keles2015}
\bibinfo{author}{\bibfnamefont{A.}~\bibnamefont{Kele\c{s}}} \bibnamefont{and}
  \bibinfo{author}{\bibfnamefont{M.~O.} \bibnamefont{Oktel}},
  \bibinfo{journal}{Phys. Rev. A} \textbf{\bibinfo{volume}{91}},
  \bibinfo{pages}{013629} (\bibinfo{year}{2015}).

\bibitem[{\citenamefont{Zhao et~al.}(2014{\natexlab{a}})\citenamefont{Zhao, Hu,
  Chang, Zhang, and Wang}}]{zhao2014a}
\bibinfo{author}{\bibfnamefont{J.}~\bibnamefont{Zhao}},
  \bibinfo{author}{\bibfnamefont{S.}~\bibnamefont{Hu}},
  \bibinfo{author}{\bibfnamefont{J.}~\bibnamefont{Chang}},
  \bibinfo{author}{\bibfnamefont{P.}~\bibnamefont{Zhang}}, \bibnamefont{and}
  \bibinfo{author}{\bibfnamefont{X.}~\bibnamefont{Wang}},
  \bibinfo{journal}{Phys. Rev. A} \textbf{\bibinfo{volume}{89}},
  \bibinfo{pages}{043611} (\bibinfo{year}{2014}{\natexlab{a}}).

\bibitem[{\citenamefont{Zhao et~al.}(2014{\natexlab{b}})\citenamefont{Zhao, Hu,
  Chang, Zheng, Zhang, and Wang}}]{zhao2014b}
\bibinfo{author}{\bibfnamefont{J.}~\bibnamefont{Zhao}},
  \bibinfo{author}{\bibfnamefont{S.}~\bibnamefont{Hu}},
  \bibinfo{author}{\bibfnamefont{J.}~\bibnamefont{Chang}},
  \bibinfo{author}{\bibfnamefont{F.}~\bibnamefont{Zheng}},
  \bibinfo{author}{\bibfnamefont{P.}~\bibnamefont{Zhang}}, \bibnamefont{and}
  \bibinfo{author}{\bibfnamefont{X.}~\bibnamefont{Wang}},
  \bibinfo{journal}{Phys. Rev. B} \textbf{\bibinfo{volume}{90}},
  \bibinfo{pages}{085117} (\bibinfo{year}{2014}{\natexlab{b}}).

\bibitem[{\citenamefont{Xu et~al.}(2014)\citenamefont{Xu, Cole, and
  Zhang}}]{xu2014}
\bibinfo{author}{\bibfnamefont{Z.}~\bibnamefont{Xu}},
  \bibinfo{author}{\bibfnamefont{W.}~\bibnamefont{Cole}}, \bibnamefont{and}
  \bibinfo{author}{\bibfnamefont{S.}~\bibnamefont{Zhang}},
  \bibinfo{journal}{Phys. Rev. A} \textbf{\bibinfo{volume}{89}},
  \bibinfo{pages}{051604(R)} (\bibinfo{year}{2014}), \eprint{arXiv:1403.3491}.

\bibitem[{\citenamefont{Peotta et~al.}(2014)\citenamefont{Peotta, Mazza,
  Vicari, Polini, Fazio, and Rossini}}]{peotta2014}
\bibinfo{author}{\bibfnamefont{S.}~\bibnamefont{Peotta}},
  \bibinfo{author}{\bibfnamefont{L.}~\bibnamefont{Mazza}},
  \bibinfo{author}{\bibfnamefont{E.}~\bibnamefont{Vicari}},
  \bibinfo{author}{\bibfnamefont{M.}~\bibnamefont{Polini}},
  \bibinfo{author}{\bibfnamefont{R.}~\bibnamefont{Fazio}}, \bibnamefont{and}
  \bibinfo{author}{\bibfnamefont{D.}~\bibnamefont{Rossini}},
  \bibinfo{journal}{J. Stat. Mech.: Theor. Exp.}
  \textbf{\bibinfo{volume}{2014}}, \bibinfo{pages}{P09005}
  (\bibinfo{year}{2014}).

\bibitem[{\citenamefont{Piraud et~al.}(2014)\citenamefont{Piraud, Cai,
  McCulloch, and Schollw{\"o}ck}}]{piraud2014}
\bibinfo{author}{\bibfnamefont{M.}~\bibnamefont{Piraud}},
  \bibinfo{author}{\bibfnamefont{Z.}~\bibnamefont{Cai}},
  \bibinfo{author}{\bibfnamefont{I.~P.} \bibnamefont{McCulloch}},
  \bibnamefont{and}
  \bibinfo{author}{\bibfnamefont{U.}~\bibnamefont{Schollw{\"o}ck}},
  \bibinfo{journal}{Phys. Rev. A} \textbf{\bibinfo{volume}{89}},
  \bibinfo{pages}{063618} (\bibinfo{year}{2014}).

\bibitem[{\citenamefont{Barbiero et~al.}(2014)\citenamefont{Barbiero, Abad, and
  Recati}}]{barbiero2014}
\bibinfo{author}{\bibfnamefont{L.}~\bibnamefont{Barbiero}},
  \bibinfo{author}{\bibfnamefont{M.}~\bibnamefont{Abad}}, \bibnamefont{and}
  \bibinfo{author}{\bibfnamefont{A.}~\bibnamefont{Recati}},
  \emph{\bibinfo{title}{Magnetic phase transition in coherently coupled bose
  gases in optical lattices}}, \bibinfo{howpublished}{arXiv:1403.4185}
  (\bibinfo{year}{2014}).

\bibitem[{\citenamefont{Fazio and van~der
  Zant}(2001)}]{fazio_josephson_junction_review}
\bibinfo{author}{\bibfnamefont{R.}~\bibnamefont{Fazio}} \bibnamefont{and}
  \bibinfo{author}{\bibfnamefont{H.}~\bibnamefont{van~der Zant}},
  \bibinfo{journal}{Phys. Rep.} \textbf{\bibinfo{volume}{355}},
  \bibinfo{pages}{235} (\bibinfo{year}{2001}).

\bibitem[{\citenamefont{Paredes et~al.}(2004)\citenamefont{Paredes, Widera,
  Murg, Mandel, Folling, Cirac, Shlyapnikov, Hansch, and
  Bloch}}]{paredes_toks_experiment}
\bibinfo{author}{\bibfnamefont{B.}~\bibnamefont{Paredes}},
  \bibinfo{author}{\bibfnamefont{A.}~\bibnamefont{Widera}},
  \bibinfo{author}{\bibfnamefont{V.}~\bibnamefont{Murg}},
  \bibinfo{author}{\bibfnamefont{O.}~\bibnamefont{Mandel}},
  \bibinfo{author}{\bibfnamefont{S.}~\bibnamefont{Folling}},
  \bibinfo{author}{\bibfnamefont{I.}~\bibnamefont{Cirac}},
  \bibinfo{author}{\bibfnamefont{G.}~\bibnamefont{Shlyapnikov}},
  \bibinfo{author}{\bibfnamefont{T.}~\bibnamefont{Hansch}}, \bibnamefont{and}
  \bibinfo{author}{\bibfnamefont{I.}~\bibnamefont{Bloch}},
  \bibinfo{journal}{Nature (London)} \textbf{\bibinfo{volume}{429}},
  \bibinfo{pages}{277} (\bibinfo{year}{2004}).

\bibitem[{\citenamefont{Kinoshita et~al.}(2004)\citenamefont{Kinoshita, Wenger,
  and Weiss}}]{kinoshita_tonks_experiment}
\bibinfo{author}{\bibfnamefont{T.}~\bibnamefont{Kinoshita}},
  \bibinfo{author}{\bibfnamefont{T.}~\bibnamefont{Wenger}}, \bibnamefont{and}
  \bibinfo{author}{\bibfnamefont{D.}~\bibnamefont{Weiss}},
  \bibinfo{journal}{Science} \textbf{\bibinfo{volume}{305}},
  \bibinfo{pages}{5687} (\bibinfo{year}{2004}).

\bibitem[{\citenamefont{Osterloh et~al.}(2005)\citenamefont{Osterloh, Baig,
  Santos, Zoller, and Lewenstein}}]{osterloh05_gauge}
\bibinfo{author}{\bibfnamefont{K.}~\bibnamefont{Osterloh}},
  \bibinfo{author}{\bibfnamefont{M.}~\bibnamefont{Baig}},
  \bibinfo{author}{\bibfnamefont{L.}~\bibnamefont{Santos}},
  \bibinfo{author}{\bibfnamefont{P.}~\bibnamefont{Zoller}}, \bibnamefont{and}
  \bibinfo{author}{\bibfnamefont{M.}~\bibnamefont{Lewenstein}},
  \bibinfo{journal}{Phys. Rev. Lett.} \textbf{\bibinfo{volume}{95}},
  \bibinfo{pages}{010403} (\bibinfo{year}{2005}).

\bibitem[{\citenamefont{Ruseckas et~al.}(2005)\citenamefont{Ruseckas,
  Juzeli\ifmmode~\bar{u}\else \={u}\fi{}nas, \"Ohberg, and
  Fleischhauer}}]{ruseckas05_gauge}
\bibinfo{author}{\bibfnamefont{J.}~\bibnamefont{Ruseckas}},
  \bibinfo{author}{\bibfnamefont{G.}~\bibnamefont{Juzeli\ifmmode~\bar{u}\else
  \={u}\fi{}nas}}, \bibinfo{author}{\bibfnamefont{P.}~\bibnamefont{\"Ohberg}},
  \bibnamefont{and}
  \bibinfo{author}{\bibfnamefont{M.}~\bibnamefont{Fleischhauer}},
  \bibinfo{journal}{Phys. Rev. Lett.} \textbf{\bibinfo{volume}{95}},
  \bibinfo{pages}{010404} (\bibinfo{year}{2005}).

\bibitem[{\citenamefont{Lin et~al.}(2011)\citenamefont{Lin, Jimenez-Garcia, and
  Spielman}}]{lin2011_soc}
\bibinfo{author}{\bibfnamefont{Y.}~\bibnamefont{Lin}},
  \bibinfo{author}{\bibfnamefont{K.}~\bibnamefont{Jimenez-Garcia}},
  \bibnamefont{and} \bibinfo{author}{\bibfnamefont{I.~B.}
  \bibnamefont{Spielman}}, \bibinfo{journal}{Nature}
  \textbf{\bibinfo{volume}{471}}, \bibinfo{pages}{83} (\bibinfo{year}{2011}).

\bibitem[{\citenamefont{Atala et~al.}(2014)\citenamefont{Atala, Aidelsburger,
  Lohse, Barreiro, Paredes, and Bloch}}]{atala2014}
\bibinfo{author}{\bibfnamefont{M.}~\bibnamefont{Atala}},
  \bibinfo{author}{\bibfnamefont{M.}~\bibnamefont{Aidelsburger}},
  \bibinfo{author}{\bibfnamefont{M.}~\bibnamefont{Lohse}},
  \bibinfo{author}{\bibfnamefont{J.}~\bibnamefont{Barreiro}},
  \bibinfo{author}{\bibfnamefont{B.}~\bibnamefont{Paredes}}, \bibnamefont{and}
  \bibinfo{author}{\bibfnamefont{I.}~\bibnamefont{Bloch}},
  \bibinfo{journal}{Nature Physics} \textbf{\bibinfo{volume}{10}},
  \bibinfo{pages}{588} (\bibinfo{year}{2014}).

\bibitem[{\citenamefont{Piraud et~al.}(2015)\citenamefont{Piraud,
  Heidrich-Meisner, McCulloch, Greschner, Vekua, and
  Schollw\"ock}}]{piraud2014b}
\bibinfo{author}{\bibfnamefont{M.}~\bibnamefont{Piraud}},
  \bibinfo{author}{\bibfnamefont{F.}~\bibnamefont{Heidrich-Meisner}},
  \bibinfo{author}{\bibfnamefont{I.~P.} \bibnamefont{McCulloch}},
  \bibinfo{author}{\bibfnamefont{S.}~\bibnamefont{Greschner}},
  \bibinfo{author}{\bibfnamefont{T.}~\bibnamefont{Vekua}}, \bibnamefont{and}
  \bibinfo{author}{\bibfnamefont{U.}~\bibnamefont{Schollw\"ock}},
  \bibinfo{journal}{Phys. Rev. B} \textbf{\bibinfo{volume}{91}},
  \bibinfo{pages}{140406} (\bibinfo{year}{2015}).

\bibitem[{\citenamefont{{Cazalilla} et~al.}(2011)\citenamefont{{Cazalilla},
  {Citro}, {Giamarchi}, {Orignac}, and {Rigol}}}]{cazalilla2011}
\bibinfo{author}{\bibfnamefont{M.~A.} \bibnamefont{{Cazalilla}}},
  \bibinfo{author}{\bibfnamefont{R.}~\bibnamefont{{Citro}}},
  \bibinfo{author}{\bibfnamefont{T.}~\bibnamefont{{Giamarchi}}},
  \bibinfo{author}{\bibfnamefont{E.}~\bibnamefont{{Orignac}}},
  \bibnamefont{and} \bibinfo{author}{\bibfnamefont{M.}~\bibnamefont{{Rigol}}},
  \bibinfo{journal}{Rev. Mod. Phys.} \textbf{\bibinfo{volume}{83}},
  \bibinfo{pages}{1405} (\bibinfo{year}{2011}),
  \bibinfo{note}{arXiv:1101.5337}.

\bibitem[{\citenamefont{{Di Dio} et~al.}(2015)\citenamefont{{Di Dio}, {De
  Palo}, Orignac, Citro, and Chiofalo}}]{supplementary}
\bibinfo{author}{\bibfnamefont{M.}~\bibnamefont{{Di Dio}}},
  \bibinfo{author}{\bibfnamefont{S.}~\bibnamefont{{De Palo}}},
  \bibinfo{author}{\bibfnamefont{E.}~\bibnamefont{Orignac}},
  \bibinfo{author}{\bibfnamefont{R.}~\bibnamefont{Citro}}, \bibnamefont{and}
  \bibinfo{author}{\bibfnamefont{M.-L.} \bibnamefont{Chiofalo}},
  \emph{\bibinfo{title}{Supplementary material for "persisting meissner state
  and \ldots"}} (\bibinfo{year}{2015}), \bibinfo{note}{see Supplemental
  Material at [URL will be inserted by publisher] for detailed bosonization
  calculation}.

\bibitem[{\citenamefont{Cr{\'e}pin et~al.}(2011)\citenamefont{Cr{\'e}pin,
  Laflorencie, Roux, and Simon}}]{crepin2011}
\bibinfo{author}{\bibfnamefont{F.}~\bibnamefont{Cr{\'e}pin}},
  \bibinfo{author}{\bibfnamefont{N.}~\bibnamefont{Laflorencie}},
  \bibinfo{author}{\bibfnamefont{G.}~\bibnamefont{Roux}}, \bibnamefont{and}
  \bibinfo{author}{\bibfnamefont{P.}~\bibnamefont{Simon}},
  \bibinfo{journal}{Phys. Rev. B} \textbf{\bibinfo{volume}{84}},
  \bibinfo{pages}{054517} (\bibinfo{year}{2011}).

\bibitem[{\citenamefont{Haldane}(1981)}]{haldane_bosons}
\bibinfo{author}{\bibfnamefont{F.~D.~M.} \bibnamefont{Haldane}},
  \bibinfo{journal}{Phys. Rev. Lett.} \textbf{\bibinfo{volume}{47}},
  \bibinfo{pages}{1840} (\bibinfo{year}{1981}).

\bibitem[{\citenamefont{White}(1992)}]{white_dmrg_letter}
\bibinfo{author}{\bibfnamefont{S.~R.} \bibnamefont{White}},
  \bibinfo{journal}{Phys. Rev. Lett.} \textbf{\bibinfo{volume}{69}},
  \bibinfo{pages}{2863} (\bibinfo{year}{1992}).

\bibitem[{\citenamefont{White}(1993)}]{white_dmrg}
\bibinfo{author}{\bibfnamefont{S.~R.} \bibnamefont{White}},
  \bibinfo{journal}{Phys. Rev. B} \textbf{\bibinfo{volume}{48}},
  \bibinfo{pages}{10345} (\bibinfo{year}{1993}).

\bibitem[{\citenamefont{Schollw{\"o}ck}(2005)}]{schollwock2005}
\bibinfo{author}{\bibfnamefont{U.}~\bibnamefont{Schollw{\"o}ck}},
  \bibinfo{journal}{Rev. Mod. Phys.} \textbf{\bibinfo{volume}{77}},
  \bibinfo{pages}{259} (\bibinfo{year}{2005}).

\bibitem[{\citenamefont{Nersesyan et~al.}(1998)\citenamefont{Nersesyan,
  Gogolin, and Essler}}]{nersesyan_incom}
\bibinfo{author}{\bibfnamefont{A.~A.} \bibnamefont{Nersesyan}},
  \bibinfo{author}{\bibfnamefont{A.~O.} \bibnamefont{Gogolin}},
  \bibnamefont{and} \bibinfo{author}{\bibfnamefont{F.~H.~L.}
  \bibnamefont{Essler}}, \bibinfo{journal}{Phys. Rev. Lett.}
  \textbf{\bibinfo{volume}{81}}, \bibinfo{pages}{910} (\bibinfo{year}{1998}).

\bibitem[{\citenamefont{Lecheminant et~al.}(2001)\citenamefont{Lecheminant,
  Jolicoeur, and Azaria}}]{lecheminant2001}
\bibinfo{author}{\bibfnamefont{P.}~\bibnamefont{Lecheminant}},
  \bibinfo{author}{\bibfnamefont{T.}~\bibnamefont{Jolicoeur}},
  \bibnamefont{and} \bibinfo{author}{\bibfnamefont{P.}~\bibnamefont{Azaria}},
  \bibinfo{journal}{Phys. Rev. B} \textbf{\bibinfo{volume}{63}},
  \bibinfo{pages}{174426} (\bibinfo{year}{2001}).

\bibitem[{\citenamefont{Jolicoeur and Lecheminant}(2002)}]{jolicoeur2002}
\bibinfo{author}{\bibfnamefont{T.}~\bibnamefont{Jolicoeur}} \bibnamefont{and}
  \bibinfo{author}{\bibfnamefont{P.}~\bibnamefont{Lecheminant}},
  \bibinfo{journal}{Prog. Theor. Phys. Supp.} \textbf{\bibinfo{volume}{145}},
  \bibinfo{pages}{23} (\bibinfo{year}{2002}).

\bibitem[{\citenamefont{Zarea et~al.}(2004)\citenamefont{Zarea, Fabrizio, and
  Nersesyan}}]{zarea04_chiral_xxz}
\bibinfo{author}{\bibfnamefont{M.}~\bibnamefont{Zarea}},
  \bibinfo{author}{\bibfnamefont{M.}~\bibnamefont{Fabrizio}}, \bibnamefont{and}
  \bibinfo{author}{\bibfnamefont{A.}~\bibnamefont{Nersesyan}},
  \bibinfo{journal}{Eur. Phys. J. B} \textbf{\bibinfo{volume}{39}},
  \bibinfo{pages}{155} (\bibinfo{year}{2004}).

\bibitem[{\citenamefont{{Greschner} et~al.}(2015)\citenamefont{{Greschner},
  {Piraud}, {Heidrich-Meisner}, {McCulloch}, {Schollw{\"o}ck}, and
  {Vekua}}}]{greschner2015}
\bibinfo{author}{\bibfnamefont{S.}~\bibnamefont{{Greschner}}},
  \bibinfo{author}{\bibfnamefont{M.}~\bibnamefont{{Piraud}}},
  \bibinfo{author}{\bibfnamefont{F.}~\bibnamefont{{Heidrich-Meisner}}},
  \bibinfo{author}{\bibfnamefont{I.~P.} \bibnamefont{{McCulloch}}},
  \bibinfo{author}{\bibfnamefont{U.}~\bibnamefont{{Schollw{\"o}ck}}},
  \bibnamefont{and} \bibinfo{author}{\bibfnamefont{T.}~\bibnamefont{{Vekua}}},
  \emph{\bibinfo{title}{{Spontaneous increase of magnetic flux and
  chiral-current reversal in bosonic ladders: Swimming against the tide}}},
  \bibinfo{howpublished}{arXiv:1504.06564} (\bibinfo{year}{2015}).

\bibitem[{\citenamefont{{Devoret} et~al.}(2004)\citenamefont{{Devoret},
  {Wallraff}, and {Martinis}}}]{devoret2004}
\bibinfo{author}{\bibfnamefont{M.~H.} \bibnamefont{{Devoret}}},
  \bibinfo{author}{\bibfnamefont{A.}~\bibnamefont{{Wallraff}}},
  \bibnamefont{and} \bibinfo{author}{\bibfnamefont{J.~M.}
  \bibnamefont{{Martinis}}}, \emph{\bibinfo{title}{{Superconducting Qubits: A
  Short Review}}}, \bibinfo{howpublished}{eprint arXiv:cond-mat/0411174}
  (\bibinfo{year}{2004}).

\bibitem[{\citenamefont{You and Nori}(2005)}]{franco2005}
\bibinfo{author}{\bibfnamefont{J.~Q.} \bibnamefont{You}} \bibnamefont{and}
  \bibinfo{author}{\bibfnamefont{F.}~\bibnamefont{Nori}},
  \bibinfo{journal}{Physics Today} \textbf{\bibinfo{volume}{58}},
  \bibinfo{pages}{42} (\bibinfo{year}{2005}).

\end{thebibliography}

\begin{thebibliography}{25}
\expandafter\ifx\csname natexlab\endcsname\relax\def\natexlab#1{#1}\fi
\expandafter\ifx\csname bibnamefont\endcsname\relax
  \def\bibnamefont#1{#1}\fi
\expandafter\ifx\csname bibfnamefont\endcsname\relax
  \def\bibfnamefont#1{#1}\fi
\expandafter\ifx\csname citenamefont\endcsname\relax
  \def\citenamefont#1{#1}\fi
\expandafter\ifx\csname url\endcsname\relax
  \def\url#1{\texttt{#1}}\fi
\expandafter\ifx\csname urlprefix\endcsname\relax\def\urlprefix{URL }\fi
\providecommand{\bibinfo}[2]{#2}
\providecommand{\eprint}[2][]{\url{#2}}

\bibitem[{\citenamefont{Jordan and Wigner}(1928)}]{jordan_transformation}
\bibinfo{author}{\bibfnamefont{P.}~\bibnamefont{Jordan}} \bibnamefont{and}
  \bibinfo{author}{\bibfnamefont{E.}~\bibnamefont{Wigner}},
  \bibinfo{journal}{Z. Phys.} \textbf{\bibinfo{volume}{47}},
  \bibinfo{pages}{631} (\bibinfo{year}{1928}).

\bibitem[{\citenamefont{Giamarchi}(2004)}]{giamarchi_book_1d}
\bibinfo{author}{\bibfnamefont{T.}~\bibnamefont{Giamarchi}},
  \emph{\bibinfo{title}{Quantum Physics in One Dimension}}
  (\bibinfo{publisher}{Oxford University Press}, \bibinfo{address}{Oxford},
  \bibinfo{year}{2004}).

\bibitem[{\citenamefont{Haldane}(1981)}]{haldane_bosons}
\bibinfo{author}{\bibfnamefont{F.~D.~M.} \bibnamefont{Haldane}},
  \bibinfo{journal}{Phys. Rev. Lett.} \textbf{\bibinfo{volume}{47}},
  \bibinfo{pages}{1840} (\bibinfo{year}{1981}).

\bibitem[{\citenamefont{Donohue and
  Giamarchi}(2001)}]{donohue_commensurate_bosonic_ladder}
\bibinfo{author}{\bibfnamefont{P.}~\bibnamefont{Donohue}} \bibnamefont{and}
  \bibinfo{author}{\bibfnamefont{T.}~\bibnamefont{Giamarchi}},
  \bibinfo{journal}{Phys. Rev. B} \textbf{\bibinfo{volume}{63}},
  \bibinfo{pages}{180508(R)} (\bibinfo{year}{2001}).

\bibitem[{\citenamefont{Cr{\'e}pin et~al.}(2011)\citenamefont{Cr{\'e}pin,
  Laflorencie, Roux, and Simon}}]{crepin2011}
\bibinfo{author}{\bibfnamefont{F.}~\bibnamefont{Cr{\'e}pin}},
  \bibinfo{author}{\bibfnamefont{N.}~\bibnamefont{Laflorencie}},
  \bibinfo{author}{\bibfnamefont{G.}~\bibnamefont{Roux}}, \bibnamefont{and}
  \bibinfo{author}{\bibfnamefont{P.}~\bibnamefont{Simon}},
  \bibinfo{journal}{Phys. Rev. B} \textbf{\bibinfo{volume}{84}},
  \bibinfo{pages}{054517} (\bibinfo{year}{2011}).

\bibitem[{\citenamefont{Zamolodchikov}(1995)}]{zamolodchikov_energy_sg}
\bibinfo{author}{\bibfnamefont{A.~B.} \bibnamefont{Zamolodchikov}},
  \bibinfo{journal}{Int. Review of Modern Physics A}
  \textbf{\bibinfo{volume}{10}}, \bibinfo{pages}{1125} (\bibinfo{year}{1995}).

\bibitem[{\citenamefont{Lukyanov and
  Zamolodchikov}(1997)}]{lukyanov_sinegordon_correlations}
\bibinfo{author}{\bibfnamefont{S.}~\bibnamefont{Lukyanov}} \bibnamefont{and}
  \bibinfo{author}{\bibfnamefont{A.~B.} \bibnamefont{Zamolodchikov}},
  \bibinfo{journal}{Nucl. Phys. B} \textbf{\bibinfo{volume}{493}},
  \bibinfo{pages}{571} (\bibinfo{year}{1997}).

\bibitem[{\citenamefont{Uhrig and Schulz}(1996)}]{uhrig_excitation_staggered}
\bibinfo{author}{\bibfnamefont{G.~S.} \bibnamefont{Uhrig}} \bibnamefont{and}
  \bibinfo{author}{\bibfnamefont{H.~J.} \bibnamefont{Schulz}},
  \bibinfo{journal}{Phys. Rev. B} \textbf{\bibinfo{volume}{54}},
  \bibinfo{pages}{R9624} (\bibinfo{year}{1996}).

\bibitem[{\citenamefont{Affleck}(1986)}]{affleck86_dimerized}
\bibinfo{author}{\bibfnamefont{I.}~\bibnamefont{Affleck}},
  \bibinfo{journal}{Nucl. Phys. B} \textbf{\bibinfo{volume}{265}},
  \bibinfo{pages}{448} (\bibinfo{year}{1986}).

\bibitem[{\citenamefont{Tsvelik}(1992)}]{tsvelik_excitation_staggered}
\bibinfo{author}{\bibfnamefont{A.~M.} \bibnamefont{Tsvelik}},
  \bibinfo{journal}{Phys. Rev. B} \textbf{\bibinfo{volume}{45}},
  \bibinfo{pages}{486} (\bibinfo{year}{1992}).

\bibitem[{\citenamefont{{Ovchinnikov}}(2004)}]{ovchinnikov2004}
\bibinfo{author}{\bibfnamefont{A.~A.} \bibnamefont{{Ovchinnikov}}},
  \bibinfo{journal}{Journal of Physics Condensed Matter}
  \textbf{\bibinfo{volume}{16}}, \bibinfo{pages}{3147} (\bibinfo{year}{2004}),
  \eprint{arXiv:math-ph/0311050}.

\bibitem[{\citenamefont{Oshikawa and
  Affleck}(1997)}]{oshikawa_cu_benzoate_short}
\bibinfo{author}{\bibfnamefont{M.}~\bibnamefont{Oshikawa}} \bibnamefont{and}
  \bibinfo{author}{\bibfnamefont{I.}~\bibnamefont{Affleck}},
  \bibinfo{journal}{Phys. Rev. Lett.} \textbf{\bibinfo{volume}{79}},
  \bibinfo{pages}{2883} (\bibinfo{year}{1997}).

\bibitem[{\citenamefont{Affleck and Oshikawa}(1999)}]{oshikawa_cu_benzoate}
\bibinfo{author}{\bibfnamefont{I.}~\bibnamefont{Affleck}} \bibnamefont{and}
  \bibinfo{author}{\bibfnamefont{M.}~\bibnamefont{Oshikawa}},
  \bibinfo{journal}{Phys. Rev. B} \textbf{\bibinfo{volume}{60}},
  \bibinfo{pages}{1039} (\bibinfo{year}{1999}), \bibinfo{note}{phys. Rev. B
  \textbf{62}, 9200(E) (2000)}.

\bibitem[{\citenamefont{Orignac and Giamarchi}(2001)}]{orignac01_meissner}
\bibinfo{author}{\bibfnamefont{E.}~\bibnamefont{Orignac}} \bibnamefont{and}
  \bibinfo{author}{\bibfnamefont{T.}~\bibnamefont{Giamarchi}},
  \bibinfo{journal}{Phys. Rev. B} \textbf{\bibinfo{volume}{64}},
  \bibinfo{pages}{144515} (\bibinfo{year}{2001}), \eprint{cond-mat/0011497}.

\bibitem[{\citenamefont{Japaridze and
  Nersesyan}(1978)}]{japaridze_cic_transition}
\bibinfo{author}{\bibfnamefont{G.~I.} \bibnamefont{Japaridze}}
  \bibnamefont{and} \bibinfo{author}{\bibfnamefont{A.~A.}
  \bibnamefont{Nersesyan}}, \bibinfo{journal}{JETP Lett.}
  \textbf{\bibinfo{volume}{27}}, \bibinfo{pages}{334} (\bibinfo{year}{1978}).

\bibitem[{\citenamefont{Pokrovsky and Talapov}(1979)}]{pokrovsky_talapov_prl}
\bibinfo{author}{\bibfnamefont{V.~L.} \bibnamefont{Pokrovsky}}
  \bibnamefont{and} \bibinfo{author}{\bibfnamefont{A.~L.}
  \bibnamefont{Talapov}}, \bibinfo{journal}{Phys. Rev. Lett.}
  \textbf{\bibinfo{volume}{42}}, \bibinfo{pages}{65} (\bibinfo{year}{1979}).

\bibitem[{\citenamefont{Schulz}(1980)}]{schulz_cic2d}
\bibinfo{author}{\bibfnamefont{H.~J.} \bibnamefont{Schulz}},
  \bibinfo{journal}{Phys. Rev. B} \textbf{\bibinfo{volume}{22}},
  \bibinfo{pages}{5274} (\bibinfo{year}{1980}).

\bibitem[{\citenamefont{Chitra and Giamarchi}(1997)}]{chitra_spinchains_field}
\bibinfo{author}{\bibfnamefont{R.}~\bibnamefont{Chitra}} \bibnamefont{and}
  \bibinfo{author}{\bibfnamefont{T.}~\bibnamefont{Giamarchi}},
  \bibinfo{journal}{Phys. Rev. B} \textbf{\bibinfo{volume}{55}},
  \bibinfo{pages}{5816} (\bibinfo{year}{1997}).

\bibitem[{\citenamefont{Tsvelik}(1995)}]{tsvelik_book}
\bibinfo{author}{\bibfnamefont{A.}~\bibnamefont{Tsvelik}},
  \emph{\bibinfo{title}{Quantum Field Theory in Condensed Matter Physics}}
  (\bibinfo{publisher}{Cambridge University Press},
  \bibinfo{address}{Cambridge}, \bibinfo{year}{1995}).

\bibitem[{\citenamefont{Nersesyan et~al.}(1998)\citenamefont{Nersesyan,
  Gogolin, and Essler}}]{nersesyan_incom}
\bibinfo{author}{\bibfnamefont{A.~A.} \bibnamefont{Nersesyan}},
  \bibinfo{author}{\bibfnamefont{A.~O.} \bibnamefont{Gogolin}},
  \bibnamefont{and} \bibinfo{author}{\bibfnamefont{F.~H.~L.}
  \bibnamefont{Essler}}, \bibinfo{journal}{Phys. Rev. Lett.}
  \textbf{\bibinfo{volume}{81}}, \bibinfo{pages}{910} (\bibinfo{year}{1998}).

\bibitem[{\citenamefont{Lecheminant et~al.}(2001)\citenamefont{Lecheminant,
  Jolicoeur, and Azaria}}]{lecheminant2001}
\bibinfo{author}{\bibfnamefont{P.}~\bibnamefont{Lecheminant}},
  \bibinfo{author}{\bibfnamefont{T.}~\bibnamefont{Jolicoeur}},
  \bibnamefont{and} \bibinfo{author}{\bibfnamefont{P.}~\bibnamefont{Azaria}},
  \bibinfo{journal}{Phys. Rev. B} \textbf{\bibinfo{volume}{63}},
  \bibinfo{pages}{174426} (\bibinfo{year}{2001}).

\bibitem[{\citenamefont{Jolicoeur and Lecheminant}(2002)}]{jolicoeur2002}
\bibinfo{author}{\bibfnamefont{T.}~\bibnamefont{Jolicoeur}} \bibnamefont{and}
  \bibinfo{author}{\bibfnamefont{P.}~\bibnamefont{Lecheminant}},
  \bibinfo{journal}{Prog. Theor. Phys. Supp.} \textbf{\bibinfo{volume}{145}},
  \bibinfo{pages}{23} (\bibinfo{year}{2002}).

\bibitem[{\citenamefont{Zarea et~al.}(2004)\citenamefont{Zarea, Fabrizio, and
  Nersesyan}}]{zarea04_chiral_xxz}
\bibinfo{author}{\bibfnamefont{M.}~\bibnamefont{Zarea}},
  \bibinfo{author}{\bibfnamefont{M.}~\bibnamefont{Fabrizio}}, \bibnamefont{and}
  \bibinfo{author}{\bibfnamefont{A.}~\bibnamefont{Nersesyan}},
  \bibinfo{journal}{Eur. Phys. J. B} \textbf{\bibinfo{volume}{39}},
  \bibinfo{pages}{155} (\bibinfo{year}{2004}).

\bibitem[{\citenamefont{Nersesyan et~al.}(1993)\citenamefont{Nersesyan, Luther,
  and Kusmartsev}}]{nersesyan_2ch}
\bibinfo{author}{\bibfnamefont{A.}~\bibnamefont{Nersesyan}},
  \bibinfo{author}{\bibfnamefont{A.}~\bibnamefont{Luther}}, \bibnamefont{and}
  \bibinfo{author}{\bibfnamefont{F.}~\bibnamefont{Kusmartsev}},
  \bibinfo{journal}{Phys. Lett. A} \textbf{\bibinfo{volume}{176}},
  \bibinfo{pages}{363} (\bibinfo{year}{1993}).

\bibitem[{\citenamefont{Itzykson and Zuber}(1980)}]{itzykson-zuber}
\bibinfo{author}{\bibfnamefont{C.}~\bibnamefont{Itzykson}} \bibnamefont{and}
  \bibinfo{author}{\bibfnamefont{J.~B.} \bibnamefont{Zuber}},
  \emph{\bibinfo{title}{Quantum Field Theory}} (\bibinfo{publisher}{Mc Graw
  Hill}, \bibinfo{address}{New-York}, \bibinfo{year}{1980}).

\end{thebibliography}

\end{document}